%% file: paper.tex
\RequirePackage{lineno}
\documentclass[twocolumn,showpacs,aps,floatfix,prd]{revtex4}

\usepackage{graphicx}
\usepackage{dcolumn}
\usepackage{amsmath}
\usepackage{epsfig}

\input babarsym.tex
\newcommand{\BABARPubYear}    {14}
\newcommand{\BABARPubNumber}  {014}

\newcommand{\SLACPubNumber} {16172}

\def\figurebox#1#2#3{%
    \def\arg{#3}%
    \ifx\arg\empty
    {\hfill\vbox{\hsize#2\hrule\hbox to #2{\vrule\hfill\vbox to #1{\hsize#2\vfill}\vrule}\hrule}\hfill}%
    \else
    {\hfill\epsfbox{#3}\hfill}%
    \fi}

\newcommand{\Do}{D^0}

\newcommand{\GeV}{\rm{GeV}}

\newcommand{\ndf}{\rm{NDF}}

\newcommand{\ba}{\begin{array}}
\newcommand{\ea}{\end{array}}
\newcommand{\bc}{\begin{center}}
\newcommand{\ec}{\end{center}}
\newcommand{\beq}{\begin{eqnarray}}
\newcommand{\eeq}{\end{eqnarray}}
\newcommand{\bes}{\begin{eqnarray*}}
\newcommand{\ees}{\end{eqnarray*}}
\newcommand{\Zz}{\ifmmode {\rm Z} \else ${\rm Z } $ \fi}
\newcommand{\xxbar}{\ifmmode {\rm x\bar{x}} \else ${\rm x\bar{x}} $ \fi}
\newcommand{\rphi}{\ifmmode {\rm R\phi} \else ${\rm R\phi} $ \fi}

\newcommand{\Ddecpi}{{D}^{0} \rightarrow \pi^- e^+ \nu_e }
\newcommand{\Bdecpi}{{B}^{0} \rightarrow \pi^- e^+ \nu_e }
\newcommand{\DdecK}{{D}^{0} \rightarrow {K}^- e^+ \nu_e }

\long\def\inst#1{\par\nobreak\kern 4pt\nobreak
  {\it #1}\par\vskip 10pt plus 3pt minus 3pt}

\begin{document}
\preprint{\babar-PUB-\BABARPubYear/\BABARPubNumber} 
\preprint{SLAC-PUB-\SLACPubNumber} 

\begin{flushleft}
\babar-PUB-\BABARPubYear/\BABARPubNumber\\
SLAC-PUB-\SLACPubNumber\\
\end{flushleft}

\renewcommand{\thefootnote}{\arabic{footnote}}
\setcounter{footnote}{0}

\title{
{\large \bf
Measurement of the \boldmath{$\Do \rightarrow \pi^- e^+ \nu_e$} differential decay branching fraction as a function of $q^2$ and study of form factor parameterizations} 
}

\input authors_jun2014.tex

\begin{abstract}
Based on a sample of 500 million $e^+e^- \to \ccbar$ events recorded by the \babar\ detector at  c.m. energies of close to 10.6 \gev, we report on a study of the decay $\Dz \rightarrow \pi^- e^+ \nu_e$. We measure the ratio of branching fractions
$R_D = {\cal B}(D^0 \rightarrow \pi^- e^+ \nu_e)/{\cal B}(D^0 \rightarrow K^- \pi^+) = 0.0713 \pm 0.0017_{stat.} \pm 0.0024_{syst.}$, 
and use the present world average for 
${\cal B}(D^0 \rightarrow K^- \pi^+)$ to obtain ${\cal B}(\Dz \rightarrow \pi^- e^+ \nu_e) = (2.770 \pm 0.068_{{\rm stat.}} \pm 0.092_{{\rm syst.}} \pm 0.037_{{\rm ext.}})\times 10^{-3}$ where the third error accounts for the uncertainty on the
branching fraction for the reference channel. 
The measured dependence of the differential branching fraction on $q^2$, the 
four-momentum transfer squared between the $D$ and the $\pi$ meson, is compared to various 
theoretical predictions for the hadronic form factor, $f_{+,D}^{\pi}(q^2)$, and the normalization
$\Vcd \times f_{+,D}^{\pi}(q^2=0)=0.1374\pm0.0038_{{\rm stat.}} \pm0.0022_{{\rm syst.}} \pm0.0009_{{\rm ext.}}$ is extracted  from a fit to data.  Using the most recent LQCD prediction of  $f_{+,D}^{\pi}(q^2=0)=0.666 \pm 0.029$, we obtain $\Vcd=0.206 \pm 0.007_{\rm exp.} \pm 0.009_{\rm LQCD}$.  Assuming instead, 
$\Vcd=\Vus=0.2252 \pm 0.0009$, we obtain $f_{+,D}^{\pi}(q^2=0)=0.610 \pm 0.020_{\rm exp.} \pm 0.005_{\rm ext.}$. The $q^2$ dependence of $f_{+,D}^{\pi}(q^2)$ is compared to a variety of multi-pole parameterizations. This information is applied to $\Bdecpi$ decays and, combined with an earlier $\Bdecpi$ measurement 
by \babar, is used to derive estimates of $\Vub$.

\end{abstract}


\maketitle


\section{Introduction}
\label{sec:Introduction}
Precision measurements of the elements of the Cabibbo-Kobayashi-Maskawa (CKM) quark-mixing matrix rely primarily on decay rate measurements of either nuclear $\beta$ decay, or leptonic and semileptonic decays of $\pi$, $K$, $D$, and $B$ mesons.  The rates for exclusive semileptonic decays of mesons are proportional to the square of the product of the specific CKM element and form factors which are introduced to account for hadronization effects.
Various Lorentz invariant form factor calculations, models, and parameterizations have been developed to describe these perturbative and non-perturbative QCD processes. Theoretical uncertainties in these form factor predictions  significantly impact the extraction of the CKM elements from semileptonic decays, in particular $\Vub$. 

In the following, we present a measurement of the $q^2$ dependence of the Cabibbo-suppressed semileptonic $\Ddecpi$ decay rate, where $q^2= (P_D - P_{\pi})^2$ refers to the four-momentum transfer squared between initial and final state meson. Charge conjugate states are implied throughout the document. This analysis exploits the large production of charm mesons via the process $e^+ e^- \to \ccbar$ and identifies $\Dz$ from the decay $\Dstarp \to \Dz \pip$.
The momentum of the signal $\Dz$ is derived from all particles reconstructed in the event. 
A very similar method was successfully employed in the \babar\ analysis of the Cabibbo-favored 
$\DdecK$~\cite{ref:kenu} decay.
The validity of this procedure is examined and the associated systematic uncertainties reduced by   
analyzing in parallel the two-body decay $\Dz \to \Km \pip$.
From the ratio of branching fractions,
$R_D = {\cal B}(D^0 \rightarrow \pi^- e^+ \nu_e)/{\cal B}(D^0 \rightarrow K^- \pi^+)$,
we derive the  absolute value of the $\Ddecpi$ branching fraction, 
using the world average for the branching fraction for the normalization,
${\cal B}(D^0 \rightarrow K^- \pi^+)$.

The $\Ddecpi$ decay rate is proportional to the square of the product  $\Vcd \times f_{+,D}^{\pi}(q^2)$ which can be extracted from the measured distribution.
$f_{+,D}^{\pi}(q^2)$ is the corresponding hadronic form factor and is
defined in Section \ref{sec:fq2}.
 Using the LQCD prediction for the form factor normalization $f_{+,D}^{\pi}(q^2=0)$, we extract 
$\Vcd$.  Alternatively, using the most precise determination of
$\Vus=0.2252 \pm 0.0009$ from kaon decays~\cite{ref:pdg2013}, and the Wolfenstein parameterization of the CKM matrix with, neglecting terms of
order $\lambda^5$, $\Vcd=\Vus=\lambda$, we determine the hadronic form factor, its normalization, and 
$q^2$ dependence. We compare the measurements with predictions of QCD calculations and various form factor 
parameterizations.
Furthermore, we follow a procedure suggested by theorists~\cite{ref:damirfdstar} to use the information extracted in terms of certain
form factor parameterizations for $\Ddecpi$ decays and adapt them to $\Bdecpi$ decays
\cite{ref:damirfdstar} to arrive at estimates for $\Vub$.

Measurements of $\DdecK$ and $\Ddecpi$ decays were first published by the CLEO \cite{ref:cleo3}, FOCUS \cite{ref:focus}, and   
Belle \cite{ref:belle} Collaborations, and more recently by the CLEO-c~\cite{ref:cleoc08,ref:cleoc09} Collaboration, exploiting the very large sample of tagged events recorded at the $\psi(3770)$ resonance. Operating in the same energy 
region, the BESIII Collaboration \cite{ref:bes3} has also distributed preliminary results in Summer 2014.


\section{Decay rate and form factors}

\subsection{Differential decay rate}

The decay amplitude for semileptonic $D$ decays to a final-state pseudoscalar meson  
can be written in terms of vector and scalar form factors, $f_{+,D}(q^2)$ and $f_{0,D}(q^2)$~\cite{ref:wirbel,ref:neubert94,ref:richman-burchat},
\begin{eqnarray}
\lefteqn{\langle \pi(P_{\pi})| \overline{d} \gamma^{\mu} c |D(P_D)\rangle =}  \nonumber \\
& & f_{+,D}^{\pi}(q^2)\left[(P_D+P_{\pi})^\mu -
  \frac{m_D^2 - m_{\pi}^2}{q^2}\,q^\mu\right] + \nonumber \\ 
& & \mbox{} f_{0,D}^{\pi}(q^2)\,\frac{m_D^2 - m_{\pi}^2}{q^2}\,q^\mu ,
\label{eq:pilnuamp}
\end{eqnarray}
\noindent
where $P_{\pi}$ and $P_{D}$ refer to the four-momenta of the final state pion and the parent $D$ meson, and $m_{\pi}$ and $m_{D}$ to their masses.
The  four-momenta of the final state anti-electron and neutrino are denoted
with  $P_{e}$ and $P_{\nu_e}$ respectively.     
The constraint $f_{+,D}^{\pi}(0)=f_{0,D}^{\pi}(0)$ avoids a singularity at $q^2=0$.
This expression can be simplified for electrons, because in the limit of  $m_e \ll m_D$ the second and third terms can be neglected.
We are left with a single form factor $f_{+,D}(q^2)$ and the differential decay rate becomes, 
\beq
 \frac{d \Gamma}{d q^2 d\cos\theta_e} = \frac{G^2_F}{32 \pi^3}\left ( \Vcd 
\times |f_{+,D}^{\pi}(q^2)| \right )^2 p^{*3}_\pi(q^2) \sin^2\theta_e.
\label{eq:diff_decay_rate}
\eeq

Since the $\Dz$ and the $\pim$ have zero spin, only the helicity zero component of the virtual $W$ contributes.
The decay rate depends on the third power of $p^*_\pi$, the pion momentum in the $\Dz$ rest frame. The rate also depends on 
$\sin^2\theta_e$, where $\theta_e$ is the angle of the positron
in the $e^+ \nu_e$ rest frame with respect to the direction of the pion 
in the $\Dz$ rest frame.
The variation of the rate with $q^2$
depends on the decay dynamics and needs to be determined experimentally.
The form factor normalization requires knowledge of the CKM element $\Vcd$.
  
For various form factor parameterizations, in particular in terms of pole contributions,
$\Ddecpi$ decays are of particular interest because the contribution from the lowest mass pole to
$f_{+,D}^{\pi}(q^2)$ can be determined using additional 
information (for instance, the value of the $\Dstarp$ intrinsic width), thereby gaining sensitivity to contributions from singularities due to higher mass states.

It has been suggested \cite{ref:isgur} that precise knowledge of the form factors in $\Ddecpi$ decays could be used to determine 
$f_{+,B}^{\pi}(q^2)$ in the high $q^2$ region for the $\Bdecpi$ decays, and thereby improve the extraction of $\Vub$.
For this application,
the $\Ddecpi$ measurements are extrapolated to larger values of $q^2$
to overlap with the $\Bdecpi$ physical region. 
Two approaches are proposed.
One is based on Lattice QCD (LQCD) calculations of the ratio 
$f_{+,B}^{\pi}(q^2)/f_{+,D}^{\pi}(q^2)$ and measurements of the differential rates for $\Ddecpi$ and $\Bdecpi$ decays. 
This method relies on the assumption that LQCD can predict the form factor ratio with higher
accuracy than individual form factors. 
The second approach relies on measured contributions of individual resonances to the $D$ form factor $f_{+,D}^{\pi}(q^2)$ and scaling laws that relate this information to the $B$ form factor $f_{+,B}^{\pi}(q^2)$ in order to extract a value of $\Vub$. The assumptions in this approach are described in \cite{ref:burdman,ref:damirfdstar}. 


\subsection{The \boldmath{\lowercase{$f_{+,\uppercase{D}}^{\pi}(q^2)$}} 
hadronic form factor}
\label{sec:fq2}

The most general expression for the form factor $f_{+,D}^{\pi}(q^2)$ 
satisfies the dispersion relation,
\beq
f_{+,D}^{\pi}(q^2) = \frac{1}{\pi} \int_{\left (m_D+m_{\pi}\right )^2}^{\infty} dt \frac{\mathcal{I}m(f_{+,D}^{\pi}(t))}{t-q^2-i\epsilon},
\label{eq:dispers1}
\eeq
Singularities of $f_{+,D}^{\pi}(t)$ in the complex $t$-plane
originate from the interaction of the $\c$ and $\d$ quarks
resulting in a series of charm vector states of different masses with $J^P=1^-$. 
The kinematic threshold is at $t_+=(m_{D}+m_{\pi})^2$.

In practice this series of poles is truncated: one, two or three poles are considered.
The lowest pole, the $D^{*+}$  is located just above threshold and its 
contribution can be isolated because of its narrow width, 
of the order $0.1~\mevcc$.
The next pole (denoted $D_1^{*\prime}$ in the following)
has a mass of $(2610 \pm 4)\, \mevcc$ and width of $(93 \pm 14)\, \mevcc$ 
and corresponds to the first radial vector excitation 
\cite{ref:dstarprime}. 
The LHCb collaboration \cite{ref:dstarprime_lhcb} has measured somewhat different
values of $(2649 \pm 5)\, \mevcc$- and 
$(140 \pm 25)\, \mevcc$ for the mass and width of this state. However,
considering other sources of uncertainties, these differences have very little
impact on the present analysis.
Since hadronic singularities
(poles and cuts) are above the physical region, it is expected
that $f_{+,D}^{\pi}(q^2)$ is a monotonically rising function of $q^2$. 

In the following, we discuss various theoretical approaches and their parameterizations 
which are used to describe the $q^2$ dependence of the $D$ meson form factor 
$f_{+,D}^{\pi}(q^2)$.

\subsubsection{Dispersive approach with constraints}

\label{sec:burdman}
Several constraints have to be verified by the 
dispersion relations for the form factor~\cite{ref:burdman}. 
These include chiral symmetry,
Heavy Quark Symmetry (HQS), and perturbative QCD as the asymptotic behavior.
Using $H$ to denote a heavy $D$ or $B$ meson, the integral 
in Eq. (\ref{eq:dispers1}) can be expressed in terms of three contributions:
\begin{itemize}
\item the $H^*$ pole contribution, which is dominant;
\item the sum of radially excited, $J^P=1^-$, resonances noted $H_i^{*\prime}$;
\item the contribution from $H\pi$ continuum.
\end{itemize}
\begin{eqnarray}
f_{+,H}^{\pi}(q^2) &=& \frac{Res(f_{+,H}^{\pi})_{H^*}}{m^2_{H^*}-q^2}
+\sum_{i}\frac{Res(f_{+,H}^{\pi})_{H_i^{*\prime}}}{m^2_{H_i^{*\prime}}-q^2}\nonumber\\
&+&\frac{1}{\pi} \int_{t_+}^{\Lambda^2} dt \frac{\mathcal{I}m(f_{+,H}^{\pi,{\rm cont.}}(t))}{t-q^2-i\epsilon}.\label{eq:dispers2}
\end{eqnarray}
In this expression, the quantities $Res(f_{+,H}^{\pi})_{H^{*(\prime)}_{(i)}}$ are the residues
for the different vector resonances $H^{*(\prime)}_{(i)}$.
The integral over the continuum is evaluated between the threshold
and the first radial excited state ($\Lambda \sim m_{H_1^{*\prime}}$).
Contributions from orbital excitations are expected to be small 
\cite{ref:burdman}.

The residue which defines the contribution of the $H^*$ resonance can
be expressed in terms of the meson decay constant $f_{H^*}$, and 
$g_{H^*H\pi}$, the coupling to the $H\pi$ final state,
\beq
Res(f_{+,H}^{\pi})_{H^*}=\frac{1}{2}m_{H^*}\left ( \frac{f_{H^*}}{f_H} \right ) f_{H}\,g_{H^*H\pi}.
\label{eq:res1} 
\eeq

Similar expressions can be derived for the higher mass states $H_i^{*\prime}$. 
The expected values for the residues at the first two poles are
given in Appendix \ref{sec:appendixb}.

Using the behavior of the form factor at very large values of $q^2$,
a constraint 
(commonly referred to as superconvergence condition)
is obtained on the residues \cite{ref:burdman},
\beq
Res(f_{+,H}^{\pi})_{H^*}+\sum_i{Res(f_{+,H}^{\pi})_{H_i^{*(\prime)}}}+c_H \simeq 0 ,
\label{eq:superconv} 
\eeq
which can be compared to measurements; $c_H$ denotes the contribution
from continuum. 

\subsubsection{Multi-pole parameterizations}

\label{sec:3poles}
Limiting the contributions to three poles, 
the following expression is obtained,
\begin{eqnarray}
f_{+,D}^{\pi}(q^2)=\frac{f_{+,D}^{\pi}(0)}{1-c_2-c_3}\left(\frac{1}{1-\frac{q^2}{m_{D^*}^2}} 
-\sum\limits_{i=2}^{3}\frac{c_i}{1-\frac{q^2}{m_{D^{*\prime}_i}^2}}\right).
\label{eq:threepoles}
\end{eqnarray}
The coefficients $c_i$ are related to the residues introduced previously
through the following expression,
$c_i=-(m^2_{D^*}/m_{D^{*\prime}_i}^2)\times(Res(f_{+,D}^{\pi})_{D_i^{*(\prime)}}/Res(f_{+,D}^{\pi})_{D^{*}})$.

The variation with $q^2$ of each component is determined by the pole masses.
In addition to the $D^*$ pole, we fix the mass
of the first radial excitation at
 $2.61\,\gevcc$ \cite{ref:dstarprime}.  For the higher radial excitation 
we either use a fixed value of  $3.1\,\gevcc$ \cite{ref:godfreyisgur} 
(fixed three-pole ansatz) or an effective pole mass corresponding to the sum of contributions from
all poles at higher masses (effective three-pole ansatz).
Values expected for the residues at the $D^*$ (Eq.(\ref{eq:res2}))
and  at the $D^{*\prime}_1$ (Eq.(\ref{eq:res4})) can be used as constraints.
In the fixed three-pole ansatz, the constraint on the value of the residue
at the  $D^{*\prime}_1$ pole is used. In the effective three-pole ansatz,
constraints at the two poles are used and the value of the residue
at the effective pole is given by the superconvergence condition 
(Eq. (\ref{eq:superconv})). These constraints are entered in the likelihood 
function by including  Gaussian distributions centered at the expected values 
with standard deviations equal to the corresponding expected uncertainties.

Given the fact that the hadronic form factor is dominated
 by the $D^*$ pole, other contributions can be accounted for
by an effective pole at higher mass, resulting in a two-pole 
ansatz~\cite{ref:bk},

\begin{equation}
f_{+,D}^{\pi}(q^2)= f_{+,D}^{\pi}(0) \frac{1-\delta_{\rm pole}\frac{q^2}{m_{D^*}^2}}{\left (1-\frac{q^2}{m_{D^*}^2}\right ) \left (1-\beta_{\rm pole} \frac{q^2}{m_{D^*}^2} \right )} ,
\label{eq:twopoles}
\end{equation}
where $f_{+,D}^{\pi}(0)$, $\delta_{\rm pole}$ and  $\beta_{\rm pole}$ are free parameters that are extracted by a fit to data. In the present analysis, the 
expected value of the residue at the $D^*$ pole is used as a constraint in the fits.

If, in addition, the form factors $f_{+,D}^{\pi}$ and $f_{0,D}^{\pi}$ meet
certain conditions, expected to be valid at large recoil in the heavy quark limit~\cite{ref:bk},
then the ansatz can be further simplified,

\beq
f_{+,D}^{\pi}(q^2) =  \frac{f_{+,D}^{\pi}(0)}{ \left (1-\frac{q^2}{m_{D^*}^2}\right ) 
\left ( 1-\alpha_{\rm pole} \frac{q^2}{ m_{D^*}^2}\right )},
\label{eq:modpolemass}
\eeq
with two free parameters $f_{+,D}^{\pi}(0)$ and $\alpha_{\rm pole}$.
This modified-pole ansatz can be further simplified, 
\beq
f_{+,D}^{\pi}(q^2)= \frac{f_{+,D}^{\pi}(0)}{1-\frac{q^2}{m_{\rm pole}^2}},
\label{eq:pole}
\eeq
where $m_{\rm pole}$ is the single free parameter.
Of course, such an effective pole mass has no clear
interpretation and the proposed $q^2$ variation does not comply 
with constraints from QCD. The obtained pole-mass value may nonetheless be 
useful for comparisons with results from different experiments. 

\subsubsection{$z$-expansion}

\label{sec:cauchy}
The $z$-expansion is a model-independent parameterization which is based on general properties of analyticity, unitarity and crossing symmetries. Except for physical poles and thresholds, form factors are analytic functions of $q^2$, and can be expressed as a convergent power series, given a change of variables~\cite{ref:beforehill,ref:bourrely,ref:lebed2,ref:lebed3,ref:lellouch,ref:arnesen} of the following form,
\beq
z(t,t_0) = \frac{\sqrt{t_+-t} - \sqrt{t_+-t_0}}{\sqrt{t_+-t} + \sqrt{t_+-t_0}} ,
\eeq 
where $t_0=t_+(1 - \sqrt{1-t_-/t_+})$
with $t_-=q^2_{{\rm max}}=(m_{D}-m_{\pi})^2\sim 2.98~GeV^2$.
This transformation maps the kinematic region for the semileptonic decay ($0<q^2<t_-$)
onto a real segment extending over the range $ |z|_{{\rm max}}=0.167$.
More details on this parameterization are given in Appendix \ref{sec:appendixa}.

In terms of the variable $z$, the form factor, 
consistent with constraints from QCD, takes the form,
\beq
f_{+,D}^{\pi}(t)=\frac{1}{P(t) \Phi(t,t_0)}\sum_{k=0}^{\infty}a_k(t_0)~ z^k(t,t_0) ,
\label{eq:taylor}
\eeq
where $P(t)=1$ and $\Phi(t,t_0)$ is an arbitrary analytical function 
for which the  \textquotedblleft standard \textquotedblright  
choice is given in  
Appendix \ref{sec:appendixa}.
The $z$-expansion provides a parameterization
within the physical region and is well suited for fits to data 
and converges readily.  
The commonly used parameters are defined as $r_k = a_k / a_0$ for $k = 1,2$,  
and the overall normalization of the expansion 
is chosen to be $\Vcd \times f_{+,D}^{\pi}(0)$.

The $z$-expansion has some disadvantages in comparison 
to phenomenological approaches~\cite{ref:alainseb}. 
Specifically, there is no simple interpretation of the coefficients $a_k(t_0)$.
The contribution from the first pole ($\Dstarp$) is difficult 
to obtain because it requires extrapolation beyond the physical region while
the other coefficients are only weakly constrained by the available data. 

\subsubsection{ISGW2 quark model}

For completeness, we also list ISGW2~\cite{ref:isgw2}, a constituent quark model 
with relativistic corrections.  Predictions are normalized at $q^2_{\rm max}=t_-$.
The form factor is parameterized as
\begin{equation}
f_{+,D}^{\pi}(q^2) = f(q^2_{\rm max}) \left( 1+ \frac{1}{12} \alpha_I (q^2_{\rm max} - q^2)\right)^{-2} ,
\end{equation}
where $\alpha_I =\xi^2/12$ and $\xi$ is the charge radius of the final-state meson.
The uncertainties of the predictions are difficult to quantify.

\subsubsection{Summary of form factor parameterizations}

The different parameterizations of $f_{+,D}^{\pi}(q^2)$ considered in 
this analysis are listed in Table \ref{tab:expect}, along with the 
parameters and constraints considered.  

{\small
\begin{table*}
  \caption {Overview of $f_{+,D}^{\pi}(q^2)$ parameterizations.
In the fixed three-pole ansatz, the value expected for
$Res(f^{\pi}_{+,D})_{D^{*\prime}_1}$ (Eq. (\ref{eq:res4})) is used as a constraint 
whereas
in the effective three-pole ansatz the values expected for the residues
at the $D^*$ (Eq. (\ref{eq:res2})) and $D^{*\prime}_1$ (Eq. (\ref{eq:res4})) poles are used as constraints and the
value of the residue at the effective pole is given by the superconvergence condition  (Eq. (\ref{eq:superconv})). In the two poles ansatz, the value expected for the residue at the 
$D^*$ pole  (Eq. (\ref{eq:res2})) is used as constraint. These constraints are entered in fits assuming
that their expected values have Gaussian distributions.}
\begin{center}
  \begin{tabular}{lll}
    \hline\hline
Ansatz & Parameters & Constraints\\
\hline
$z$-expansion \cite{ref:beforehill} & $a_0,~r_k=a_k/a_0$ &  \\
effective three-pole& $Res(f^{\pi}_{+,D})_{D^*},~Res(f^{\pi}_{+,D})_{D^{*\prime}_1},~m_{\rm pole3}$   & $Res(f^{\pi}_{+,D})_{D^*},~Res(f^{\pi}_{+,D})_{D^{*\prime}_1}$ \\
fixed three-pole& $f_{+,D}(0),~c_2,~c_3$   & $Res(f^{\pi}_{+,D})_{D^{*\prime}_1}$ \\
two poles \cite{ref:bk}& $f_{+,D}(0),~\beta_{\rm pole},~\delta_{\rm pole}$   & $Res(f^{\pi}_{+,D})_{D^*}$\\
modified pole \cite{ref:bk}&$f_{+,D}(0),~\alpha_{\rm pole}$ & \\
simple pole &$f_{+,D}(0),~m_{\rm pole}$  & \\
ISGW2 \cite{ref:isgw2} & $f_{+,D}(t_-),~\alpha_I$ & \\
\hline\hline
  \end{tabular}
\end{center}
\label{tab:expect}
\end{table*}
} 

\subsection{Comparison of $f_{+,D}^{\pi}(q^2)$ and $f_{+,B}^{\pi}(q^2)$}

Form factor studies for $\Ddecpi$ decays are of particular interest because 
LQCD calculations are expected to result in predictions for the ratio 
of hadronic form factors for $B$ and $D$ mesons with a better accuracy 
than for the form factors of the individual mesons.

Two independent approaches to predict $f_{+,B}^{\pi}(q^2)$ based on $f_{+,D}^{\pi}(q^2)$
are considered (see Section \ref{sec:dtob}),
\begin{itemize}
\item fits to $f_{+,D}^{\pi}(q^2)$ according to the fixed three-pole ansatz as specified in
Eq. (\ref{eq:threepoles}) are used
to estimate the variation of ${\cal B}(\Bdecpi)$ as a function of the pion energy,
under the assumption that the ratio of the hadronic
form factors in $B$ and $D$ decays is largely insensitive to the energy of the final state pion;

\item using the effective three-poles ansatz given in 
Eq (\ref{eq:threepolesb}), with the value of the residue at the 
$B^*$ pole obtained from LQCD, and imposing the superconvergence condition.
\end{itemize}

Though estimates for the form factor ratios are not yet available, we 
discuss some aspects in Appendix \ref{sec:appendixc} which indicate 
that this approach may be promising in the future for larger data samples.


\section{The \babar\ Detector and Data Sets}
\label{sec:babar}

\subsection{Detector}

A detailed description of the \babar\ detector and the algorithms used
for charged and neutral particle reconstruction and identification is 
provided elsewhere~\cite{ref:babar, ref:babardet}. 
Charged particles are reconstructed by matching hits in 
the 5-layer silicon vertex tracker (SVT) 
with track elements in the 40-layer drift chamber (DCH), 
filled with a gas mixture of helium and isobutane.
Particles of low transverse momentum with an insufficient
number of DCH hits  
are reconstructed in the SVT.
Charged hadron identification is performed combining the measured 
ionization losses in the SVT and in the DCH with the information from the
Cherenkov detector (DIRC). 
Electrons are identified by the ratio of the track momentum to the
associated energy in the CsI(Tl) electromagnetic calorimeter (EMC), the transverse profile of the shower, the ionization loss in the DCH, and the Cherenkov angle in the DIRC.
Photon energies are measured in the EMC. 

\subsection{Data and MC Samples}
The data used in this analysis were recorded with the 
\babar\ detector at the \pep2 energy-asymmetric $\epem$ collider.
The results presented here were obtained using
$\epem \rightarrow \ccbar$ events from a sample with 
a total integrated luminosity of $347.2~\fb^{-1}$ \cite{ref:lumipaper}, collected
at the \FourS\ resonance (on-peak data) at $10.58 \gev$ center-of-mass (c.m.) energy.
An additional sample of $36.6~\fb^{-1}$ was recorded $40 \mev$ below
(off-peak data), just below the threshold for \BB\ production.

The normalization of off-peak and on-peak data samples
is derived from luminosity measurements, which are based on the number of detected $\mup \mun$ pairs and the QED cross section for 
$\epem \to \mup \mun (\gamma)$ production.

At $10.6 \gev$ c.m. energy, the non-resonant cross section for $\epem \to \qqbar$ with $q=(u,d,s,c)$ (referred to as continuum) is 3.4\nb , compared to the \FourS\ peak cross section of 1.05\nb . 
We use Monte Carlo (MC) techniques~\cite{ref:evtgen} to simulate the production and decay of \BB\ and \qqbar\ pairs and the detector 
response~\cite{ref:geant4}. The quark fragmentation in continuum events is simulated using JETSET~\cite{ref:jetset}.  The MC simulations include radiative effects, such as bremsstrahlung
in the detector material and initial-state and final-state
radiation~\cite{ref:photos}.

The size of the Monte Carlo (MC) event samples for $\FourS$ decays, $\ccbar$ pairs, and light quark pairs from continuum, exceed the data samples by factors of $3.3,~1.7~{\rm and}~1.1$, respectively.
These simulated samples are primarily used to study the background composition
and suppression.
Dedicated samples of nine times the size of the data sample
of pure signal events, i.e., $\ccbar$ events with 
$\Dstarp \to \Dz \pi^+_s$ decay, followed by the signal $\Ddecpi$ decay,
were generated and used to account for efficiencies and resolution effects. 
These samples were generated using the modified
pole parameterization
for $f_{+,D}(q^2)$ with $\alpha_{\rm pole}^{\pi}=0.44$ as defined 
in Eq. (\ref{eq:modpolemass}).

The MC distributions are normalized to the 
data luminosity, using the following cross sections: 
1.3~nb for $\ccbar$, 0.525~nb for $\Bp\Bm$ and $\BzBzb$, and 2.09 nb for light $\uubar,\ddbar, \ssbar$ quark pairs.


\section{Signal reconstruction}

\label{sec:Analysis}
We reconstruct signal $\Dz \rightarrow \pim \ep \nue (\g)$ decays,
in events produced in $\epem$ annihilation to $\ccbar$,
with the $\Dz$ originating from a $\Dstarp \to \Dz \pi^+_s$ decay.
The decay channel includes photons from final state radiation.  

In parallel, we reconstruct the reference sample of $\Dz \to \Km \pip (\g)$
decays,
with the $\Dz$ also originating from a $\Dstarp$ decay.  This sample
has the same number of final state particles, except for the undetected 
neutrino. 
The data reference sample combined with the corresponding MC sample is critical for tuning  details of the $\c$ quark fragmentation and the kinematics of particles accompanying the $\Dstarp$.
Both data and MC reference samples are also used to study the reconstruction of the missing
neutrino.

This analysis follows very closely the measurement of $\DdecK$ decays in 
\cite{ref:kenu}. 
The main differences in the selection are tighter identification criteria 
on the pion candidate, a veto 
against kaons, and the use of sideband regions in the $\Delta(m)= m(\Dz \pi^+_s)- m(\Dz)$ mass distribution 
to assess the different combinatorial and peaking background contributions.

In the following, we present the principal features of this
analysis, emphasizing those that differ from the previous analysis.

\subsection{Signal Selection}
\label{sec:Analysisq2}

This analysis exploits the two-jet topology of $\epem \to \ccbar$
events, generated by the largely independent, hard fragmentation of the two c-quarks. 
We divide the event into two hemispheres.  For this purpose,
all charged and neutral particle momenta are measured in the c.m. system, and 
a common thrust axis is determined. The plane which crosses the interaction point and is perpendicular to the thrust axis defines the two hemispheres.  To improve the event containment, only events with a polar angle of the thrust axis in the range $|\cos(\theta_{{\rm thrust}})|<0.6$ are retained.

In each hemisphere, we search for a positron and pion of opposite charge, and require that the positron (or electron for the charge conjugate $\Dzb$ decays) has a minimum c.m. momentum of 0.5 $\gevc$. The combinatorial background level is higher in this analysis than
in the $\DdecK$ analysis because the Cabibbo-suppressed decay results 
in a final-state charged pion in place of a charged kaon. 
To reduce the contamination from $\DdecK$ decays, two cases are considered.
To avoid the presence of a charged kaon as pion candidate 
the particle identification criterion (tight identification) 
is chosen to limit the contamination from kaons to $0.4\%$. 
If the charged kaon is not the pion candidate, 
a different criterion (loose identification) is chosen to veto kaons accompanying 
the $D^0$ candidate.
In this case,
kaon candidates are identified by the condition $L_K/(L_K+L_{\pi})>0.82$,
where $L_{K}$ and $L_{\pi}$ correspond to the likelihoods for the kaon and pion hypotheses, respectively. This selection has an efficiency of $90\%$ for real
 kaons whereas pions have a probability to be signed as kaons varying between
$2.5\%$ at $2 \gevc$ and $15\%$ at $5 \gevc$.

The $\nue$ momentum is unmeasured and two kinematic fits are performed, imposing in turn the
$\Dz$ and $\Dstarp$ mass constraint. First, 
the $\Dz$ direction and the neutrino energy are estimated from all particles measured in the event. The $\Dz$ direction is taken to be opposite to the sum of the momenta of all reconstructed particles in the event, except for the pion and the positron associated with the signal candidate. 
The neutrino energy is estimated as the difference between the total energy of the hemisphere
and the sum of the energies of all reconstructed particles in this hemisphere. 
A correction, which depends on the value of the missing energy measured in the opposite hemisphere,
is applied to account for the presence of missing energy due to 
particles escaping detection, even in the absence of a neutrino from the $D^0$ decay.
The energy of each hemisphere is given by the fact that the
total event energy is divided between two objects with masses equal
to the measured hemisphere masses. 
The $\Dz$ candidate is retained
if the $\chi^2$ probability, $P(\chi^2)$, of the first kinematic fit exceeds 10$^{-2}$. Detector performance for the reconstruction of the $\Dz$ momentum and energy are derived from the  $\Dz \to \Km \pip$ reference sample.
Corrections are applied to account for observed differences between data and simulation.
Each $\Dz$ candidate is combined with a low-momentum charged pion $\pi_s^+$  of the  same charge as the lepton, in the same hemisphere.
The invariant mass of this system allows to 
measure the mass difference $\Delta(m) = m(\Dz \pi^+_s)-m(\Dz)$ and to define  a signal region as $\Delta(m)< 0.155 \gevcc$, and two sidebands as $0.155 < \Delta(m) <0.20 \gevcc$ and $\Delta(m) >0.20 \gevcc$.
The second kinematic fit constrains the invariant mass of the 
candidate $\pim \ep \nue \pi^+_s$ to fixed values.
For events in the signal region, the $\Dstarp$ mass is used whereas 
in sidebands several values differing by $0.02\,\gevcc$ are taken. 
A requirement that $P(\chi^2) > 0.01$ leads to a 
reduction of combinatorial background.
With this procedure, large samples of sideband events are kept. 


\subsection{Background Rejection}
\label{sec:backgrej}
Background events arise from $\FourS \to \BB$ decays and $\epem \to \qqbar$ continuum events.  These backgrounds are significantly reduced by 
multi-variate analyses employing two Fisher discriminants. 

\begin{figure*}[!htb]
  \begin{center}
    \mbox{\epsfig{file=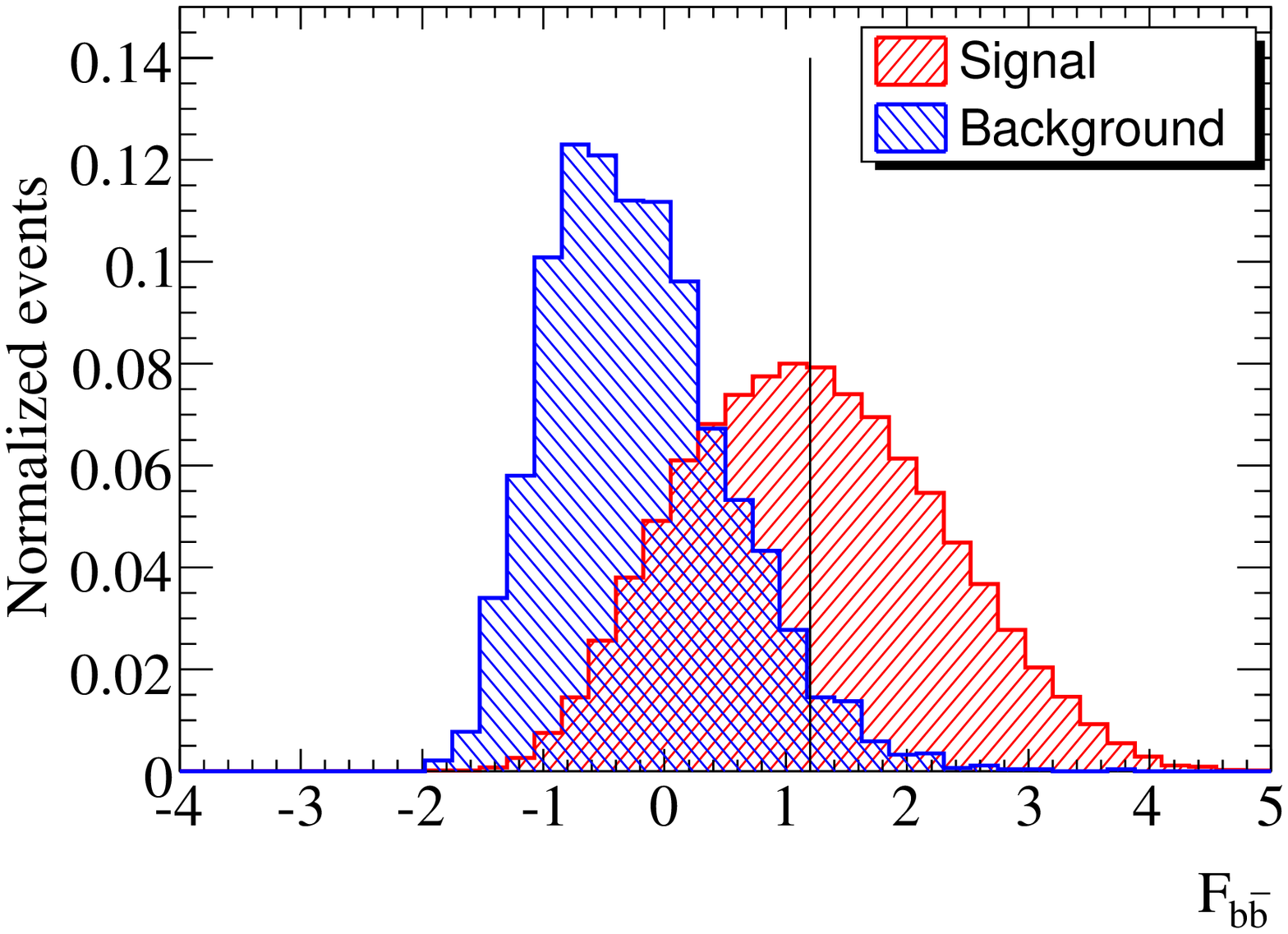,width=0.5\textwidth}
\epsfig{file=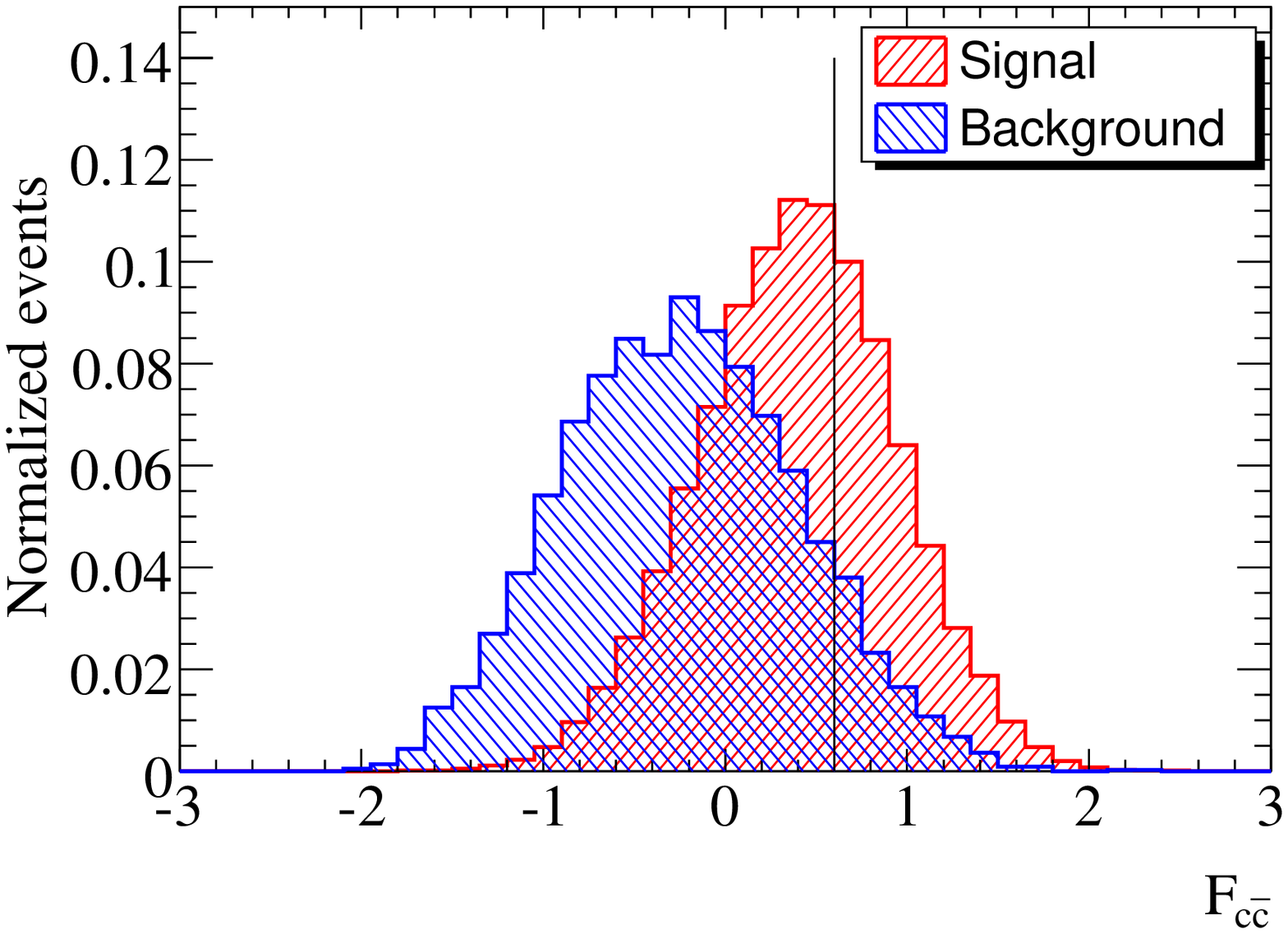,width=0.5\textwidth}}
  \end{center}
  \caption[]{ Distributions of the Fisher discriminants; Left: 
   $F_{\bbbar}$ for signal and $\BB$ events, 
   Right: $F_{\ccbar}$ for signal and other $\ccbar$ events. 
   The vertical lines indicate the selection requirements: $F_{\bbbar} > 1.2$
   and $F_{\ccbar} > 0.6$.
}
\label{fig:Fisher_bb}
\end{figure*}

To reduce the $\BB$ background, a Fisher discriminant $F_{\bbbar}$ is defined based on three variables exploiting the difference in topology of $\BB$ events and $c\bar{c}$ continuum:
\begin{itemize}
\item $R_2$, the ratio between the second and zeroth
order Fox-Wolfram moments \cite{ref:r2};
\item the total multiplicity of the detected charged and neutral particles; 
\item the momentum of the $\pi^+_s$ from the $\Dstarp \to \Dz \pi^+_s$ decay. 
\end{itemize}
 
The particle distribution in $\FourS$ events 
tends to be isotropic because the $B$ mesons are produced near threshold, while the 
particle distribution in $\ccbar$ events 
is jet-like due to the hard fragmentation of the high momentum $\c$ quarks. 
For the same reason, the 
$\Dstarp$ momenta in $\FourS$ decays are lower than in $\ccbar$ events.
The three variables are combined linearly in a Fisher discriminant. Only events with $F_{\bbbar}>1.2$  are retained.

Because few electrons are produced in 
light-quark fragmentation and lower mass particle decays, the background from the continuum arises primarily from decay of charmed particles in $\ccbar$ events.
Furthermore, the hard fragmentation function of $\c$ quarks results in charm particles and in their decay products with higher average energies and smaller angular spread (relative to the thrust axis or to the $D$ direction) compared with other particles in the hemisphere. These other particles are referred to as ``spectators'', the spectator with highest momentum 
is referred to as the ``leading'' particle. 
To reduce background from $\ccbar$ events, a Fisher discriminant 
$F_{\ccbar}$ is defined based on the same variables used in the earlier 
$\DdecK$ measurement:
\begin{itemize}
\item the $D$ momentum;
\item the invariant mass of spectators;
\item the direction of the sum of the momenta of the spectators
      relative to the thrust axis;
\item the magnitude of the momentum of the leading spectator;
\item the direction of the leading spectator relative to the $D^0$ direction;
\item the direction of the leading spectator relative to the thrust axis;
\item the direction of the lepton relative to the pion direction, in the 
$(\ep, \nu_e)$ rest frame;
\item the charged lepton momentum ($p_e$) in the c.m. frame.
\end{itemize}
The first six variables are sensitive to the properties of
$\c$ quark hadronization whereas the last two are related to the decay
characteristics of the signal decay. 
In the following, the combination of the first six variables is referred to as $F_{\ccbar -2}$.
All eight variable are combined linearly into the Fisher discriminant $F_{\ccbar}$. Only events with $F_{\ccbar}>0.6$  are retained.
Other selection requirements on $F_{\bbbar}$ and $F_{\ccbar}$ have been studied and we have used
those which correspond to the smaller systematic uncertainty for a similar total
error on fitted quantities.
Figure\,\ref{fig:Fisher_bb} shows the distribution of the two Fisher discriminants 
for the signal and background samples.

\begin{figure}[!htb]
  \begin{center}
\includegraphics[height=10cm]{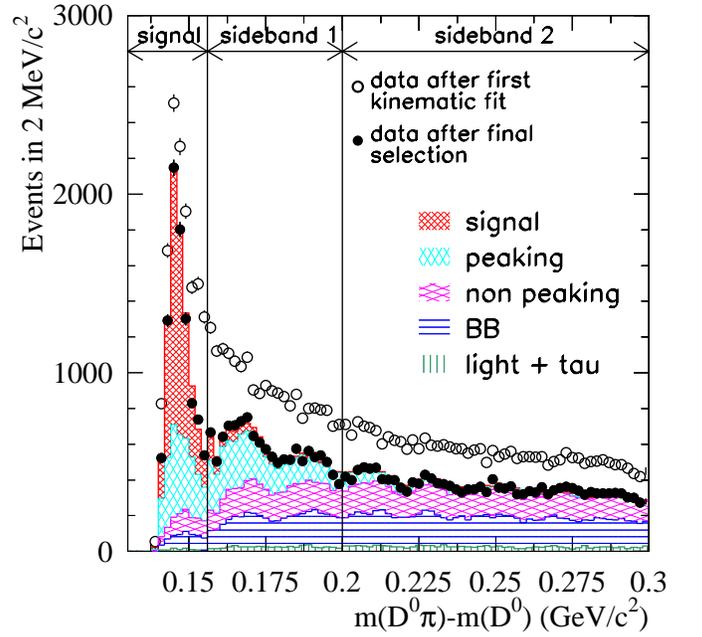}
  \caption[]{ Mass difference $\Delta(m) = m(\Dz \pi^+_s)-m(\Dz)$ after all selections criteria and the additional requirement on the first (open circles) and  second (full circles) kinematic fits probabilities.
The distribution for MC-simulated signal
and the different background distributions are superimposed for the final selections. 
These MC distributions are normalized to data based on the integrated
luminosity and  have been corrected to account for small differences  between data and MC distributions.
}
\label{fig:deltam}
\end{center}
\end{figure}

Figure \ref{fig:deltam} shows the mass difference
$\Delta(m)$ for events passing all selections criteria described above, after the sequential background suppression by the two kinematic fits.
The distributions show the expected narrow enhancement for the signal
at low $\Delta(m)$, and the suppression of the background, primarily combinatorial in nature, by the second kinematic fit. To perform detailed studies of the peaking and the non-peaking backgrounds, we use the two sidebands.

The remaining background from $\ccbar$-events can be divided into a peaking component 
 at low $\Delta(m)$ and a non-peaking component extending to higher values of $\Delta(m)$. In the signal region, the latter component
amounts to 23$\%$ of the charm background.
Peaking background events are from real $\Dstarp$ decays in which the 
slow $\pi^+_s$ is
included in the candidate track combination. 
Backgrounds from $\epem$ annihilations into light $\ddbar,\,\uubar,\, \ssbar$ pairs, $\tau^+\tau^-$ pairs and $\BB$ events are non-peaking components.

To improve the background  simulation,  simulated background distributions are corrected for observed differences between 
data and MC simulations for sideband events. Most important among them is the 
two-dimensional distribution of the $\pip$ momentum versus the missing energy in the signal hemisphere. These last corrections 
are discussed in Section \ref{sec:s4to8}.
 As a result, the  measured $\Delta(m)$ distribution is well
reproduced by the simulation and the systematic uncertainties of the signal yields are significantly reduced (for further details 
see Section~\ref{sec:Systematics}).

\begin{figure}[!htb]
\begin{center}
\includegraphics[height=9cm]{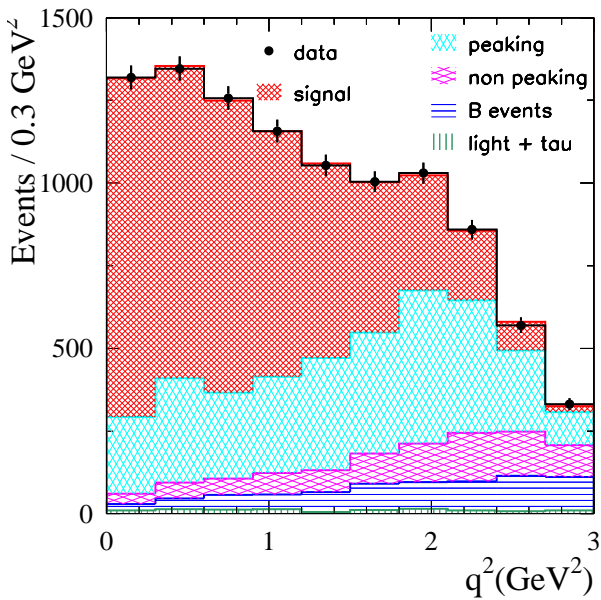}
\caption{ The measured $q^2$ distribution (data points) 
for events selected in the $\Delta(m)$ signal region is compared to the sum of 
the estimated backgrounds and the fitted signal components. Peaking and 
non-peaking background contributions refer only to $\ccbar$ events.}
\label{fig:q2rall}
\end{center}
\end{figure}

The fraction of signal events is determined from the measured 
$\Delta(m)$ distribution as the excess of events above the 
sum of the corrected background distributions.
Figure \ref{fig:q2rall} shows the $q^2= \left ( p_D-p_{\pi} \right )^2$
distribution for events selected in the signal region.
There are 9,926 signal candidates containing an estimated number
of 4,536 background events.
The selection efficiency as a function of $q^2$
varies linearly, decreasing from 1.6$\%$ at low $q^2$ to 1.0$\%$ at high $q^2$.

To obtain the true $q^2$ distribution for signal events, the 
background-subtracted
measured distribution is unfolded to correct for selection efficiency
and resolution effects.  We adopt the procedure employed in the $\DdecK$ 
analysis~\cite{ref:kenu}, and use Singular Value Decomposition (SVD) 
\cite{ref:svd} of the resolution matrix, keeping seven significant
singular values.
Table \ref{tab:q2_unfolded_tight} lists the number of selected events, the estimated total background, and the unfolded signal event yields.

\begin{table}[!htbp]
\begin{center}
  \caption[]{ {Measured number of events in bins of $q^2$: candidate events in data, estimated 
background events, and signal events corrected 
for resolution and efficiency. The first uncertainties are statistical, the second systematic,
not including those correlated with the $\Dz \to \Km \pip$ normalization sample.Because of correlations (see Table \ref{tab:errmeas}), quoted uncertainties for the total number of events differ from the values obtained when assuming uncorrelated measurements in each $q^2$ bin.}
  \label{tab:q2_unfolded_tight}}
\begin{tabular}{lcccc}
 \hline \hline
$q^2$ bin & measured & total &unfolded signal \\
  $(\gev^2)$        & events   & background & in $10^3$ unit \\
\hline
$[0.0,~0.3]  $ & $1,319$& $293 \pm 17$ & $68.3 \pm 3.5 \pm 1.2$\\
$[0.3,~0.6]$ & $1,346$& $409 \pm 21$ & $63.3 \pm 4.3 \pm 2.0$\\
$[0.6,~0.9]$ & $1,257$& $366 \pm 19$ & $61.7 \pm 3.9 \pm 1.4$\\ 
$[0.9,~1.2]$ & $1,157$& $414 \pm 21$ & $51.9 \pm 3.8 \pm 1.3$\\
$[1.2,~1.5]$ & $1,053$& $471 \pm 19$ & $41.2 \pm 3.6 \pm 1.2$\\
$[1.5,~1.8]$ & $1,004$& $548 \pm 22$ & $36.1 \pm 3.4 \pm 1.5$\\
$[1.8,~2.1]$ & $1,030$& $675 \pm 29$ & $28.6 \pm 3.2 \pm 2.4$\\
$[2.1,~2.4]$ & $859  $& $645 \pm 25$ & $16.7 \pm 2.7 \pm 2.1$\\
$[2.4,~2.7]$ & $570  $& $494 \pm 21$ & $6.5  \pm 2.4 \pm 1.2$\\
$[2.7,~q^2_{\rm max.}]$ & $331  $& $307 \pm 18$ & $1.2  \pm 0.8 \pm 0.3$\\
\hline
Total        & $9,926$& $  4,623   $ & $375.4\pm 9.2 \pm 10.1 $\\

\hline \hline
\end{tabular}
\end{center}
\end{table}


\section{Systematic Uncertainties}
\label{sec:Systematics}

Systematic uncertainties in the total branching fraction and differential decay
 rates are expected to originate from 
imperfect simulation of $\c$ quark fragmentation and of the detector response, from 
uncertainties in the background composition and the size of their contributions to the 
selected sample, and from the uncertainty in the modeling of the signal decay.
We study the origin and size of various systematic 
effects, correct the MC simulation, if possible, and assess the impact of the uncertainty in the correction of the signal distributions.
Many of these studies make use of standard \babar\ measurements of detection efficiencies, 
others rely on data control samples, and the sample of
$\Dz \rightarrow \Km \pip$ decays.
In the following study of various form factor parameterizations, we adopt 
the observed changes as contributions to the systematic uncertainties. 

A list of the systematic uncertainties from the different sources  (S1 to S20) in terms of 
variations in the numbers of unfolded signal events in each of the ten $q^2$ intervals
is presented in Table \,\ref{tab:syst1_tight_unfold}. 
The total systematic uncertainty for each interval is derived assuming no correlations among the different sources. 

\begin{table*}[!htbp]
\begin{center}
  \caption[]{ {Expected variations of the unfolded number of events in each $q^2$ interval
from the different sources of systematic uncertainties. The sign indicates whether the corresponding correction increases or decreases the signal yield. For the sources
S2, S18, and S20, these variations include only the impact on the $q^2$ variation.  The total systematic uncertainty for each interval is derived assuming no correlations among the different sources. }
  \label{tab:syst1_tight_unfold}}
\begin{tabular}{l r r r r r r r r r r }
 \hline \hline
$q^2$ bin $(\gev^2)$& [0.0, 0.3] &[0.3, 0.6] &[0.6, 0.9] &[0.9, 1.2] &[1.2, 1.5] &[1.5, 1.8] &[1.8, 2.1] &[2.1, 2.4] &[2.4, 2.7] &[2.7, $q^2_{\rm max}$] \\
\hline
S1  &  $  -360$ & $  -422$ & $  -143$ & $   260$ & $   120$ & 
  $   464$ & $  1491$ & $  1347$ & $   463$ & $    52$\\
\hline
S2 &   $   292$ & $   147$ & $  -150$ & $  -188$ & $    59$ & 
  $   -50$ & $  -144$ & $   -38$ & $    54$ & $    19$\\ 
\hline
S3  &   $   181$ & $   621$ & $   480$ & $   -84$ & $   423$ & 
  $   117$ & $  -270$ & $    100$ & $   673$ & $   248$\\
\hline
S4 &   $   309$ & $   756$ & $   496$ & $   578$ & $   859$ & 
  $  1125$ & $  1539$ & $  1288$ & $   725$ & $   194$\\ 
S5 &   $     1$ & $    -2$ & $    -1$ & $    11$ & $     9$ & 
  $     4$ & $    25$ & $    32$ & $    30$ & $     9$\\
S6 &   $  -625$ & $  -834$ & $  -536$ & $  -729$ & $  -423$ & 
  $   -88$ & $   -39$ & $   -50$ & $   -33$ & $    -7$\\ 
S7 & $ 390$ & $ 926$ & $ 294$ & $ -24$ & $ 368$ & $ 326$ &
  $ 359$ & $ 231$ & $ 222$ & $ 74$\\
S8 &   $  -137$ & $  -208$ & $     4$ & $   -48$ & $    -7$ & 
  $   -93$ & $    48$ & $    -9$ & $  -101$ & $   -31$ \\
\hline
S9  & $   -61$ & $   128$ & $    75$ & $  -108$ & $   -88$ & 
  $   -30$ & $    -10$ & $    32$ & $   -62$ & $   -18$\\
S10  &  $   -12$ & $   150$ & $   296$ & $  -331$ & $   -71$ & 
  $   385$ & $   414$ & $   346$ & $   139$ & $    26$\\
S11  &   $    54$ & $   102$ & $    56$ & $    46$ & $    30$ & 
  $    22$ & $   -21$ & $   -69$ & $   -42$ & $    -9$\\
S12  &   $   -21$ & $   114$ & $   203$ & $  -221$ & $   -33$ & 
  $   233$ & $   304$ & $   337$ & $   166$ & $    38$\\
S13  &   $    27$ & $   191$ & $   132$ & $   -50$ & $   -70$ & 
  $    12$ & $   -41$ & $  -147$ & $  -184$ & $   -50$\\
S14  &   $    94$ & $   488$ & $   186$ & $  -443$ & $     4$ & 
  $    99$ & $   324$ & $   522$ & $    81$ & $     1$\\
\hline
S15  &   $  -334$ & $   768$ & $    94$ & $  -433$ & $   -34$ & 
  $   -21$ & $    11$ & $    84$ & $   -30$ & $   -11$\\
\hline
S16 &   $  -354$ & $  -149$ & $    96$ & $  -165$ & $   196$ & 
  $   -79$ & $    81$ & $    97$ & $    34$ & $     3$ \\ 
\hline
S17  &   $   151$ & $   478$ & $   940$ & $  -122$ & $   -15$ & 
  $   492$ & $   663$ & $   442$ & $   149$ & $    22$\\
\hline
S18  &   $  -143$ & $  -157$ & $   -54$ & $    11$ & $    36$ & 
  $    81$ & $   117$ & $ 72$ & $ 29$ & $     7$\\
\hline
S19  &   $  -560$ & $  -352$ & $  -123$ & $    39$ & $   162$ & 
  $   259$ & $   282$ & $   220$ & $   116$ & $    24$ \\
\hline
S20  &   $   -46$ & $    39$ & $    96$ & $   -96$ & $   -27$ & 
  $   -78$ & $    -3$ & $    55$ & $    45$ & $    14$ \\
\hline
Total &   $1232$ & $2020 $ & $1418 $ & $1261 $ & $1156 $ & $1471$ & $2394$ & $2084$ & $1180$ & $340$ \\
\hline \hline
\end{tabular}

\end{center}
\end{table*}

\subsubsection{Charmed Meson Background (S1)}
Corrections are applied to improve the agreement between data and
MC for event samples containing an exclusively reconstructed
decay of  $\Dz$, $\Dp$, $\Ds$, or $\Dstarp$ mesons, based on a procedure 
that had previously been used in measurements of semileptonic decays 
of charm mesons~\cite{ref:kenu,ref:dskkenu,ref:dkpienu}.
We correct the simulation to match the data and,
from the measured reduction of initial differences, 
we adopt a systematic uncertainty of typically 30\% of the impact 
of the corrections on the signal yield.

\subsubsection{$\Dstarp$ Production (S2)}

To verify the simulation of $\Dz$ meson production via 
$\c$ quark fragmentation, we compare   
distributions of the variables entering in the definition of the Fisher
discriminants  $F_{\bbbar}$ and $F_{\ccbar-2}$ in data and MC  samples of 
$\Dstarp \rightarrow \Dz \pi^+_s;~\Dz \rightarrow \Km \pip$ events.
We correct the simulation of the fragmentation process
and, from the measured reduction of the differences,  take as an estimate of 
the systematic uncertainty 30\% of the observed
change in the $q^2$ distribution.  
Effects of this correction to the $\Dstarp$ production
on the measurement of $R_D$, the ratio of branching fractions 
for the two $\Dz$ decays, must be evaluated in a correlated way
for $\Dz \rightarrow \Km \pip$ and $\Ddecpi$ decays.
This is included in systematic uncertainties given in Table \ref{tab:errmeas}.
Therefore, in Table \ref{tab:syst1_tight_unfold}, we do 
not include the uncertainty due to this correction in the total number 
of fitted signal events.

\subsubsection{$\BB$ Production (S3)}
Differences in the simulation and data for $\FourS \to \BB$ decays  
are accessed by comparisons of various distributions characterizing 
$\BB$ events. To determine these differences, 
off-peak data are subtracted from on-peak data with appropriate normalization.  
The full change of the signal yield measured when
using these corrections is taken as the systematic uncertainty. 

The normalization of the $\BB$ background is fitted using events in the two sideband regions
and the corresponding uncertainty is included in the S4 systematic uncertainty. 

\subsubsection{Additional Corrections for Backgrounds (S4 - S8)}
\label{sec:s4to8}
Beyond the uncertainties in the non-peaking charm background (S1),
in the fragmentation of $\c$ quarks to produce $\Dstarp$ (S2),
and in the $\FourS \to \BB$ background (S3) that have been assessed so far,
it is important to examine additional corrections to
light-quark continuum production and the peaking and 
non-peaking charm backgrounds.

For this purpose, two-dimensional distributions of the pion momentum 
versus the missing energy in the signal 
hemisphere are examined for sideband events 
selected in off-peak and on-peak data.
The distributions are fitted to determine 15 scale factors. 
Six scale factors are adjusted for the light-quark continuum, one for each 
interval in the $\pi^-$ momentum. Six additional parameters are fitted
to scale the non-peaking charm background, for the same six $\pi^-$ momentum 
intervals. 

Five event categories are defined for the charm peaking 
background, corresponding to different distributions of the missing energy:
\begin{itemize}
\item[{\it cat 1:}] $\Dz \rightarrow \Kz \pim \ep \nue$ decays; 
\item[{\it cat 2:}] $\Dz \rightarrow \piz \pim \ep \nue$ decays; 
\item[{\it cat 3:}] the candidate pion comes from fragmentation; 
\item[{\it cat 4:}] most of these events ($>80\%$) are $\Dz \rightarrow \Km(\piz) \ep \nue$ decays with the $\Km$ identified as a tight pion.
The remaining fraction contains $\Dz \rightarrow  \pim (\piz) \ep \nue$ 
decays with the candidate $\pim$ coming from the other $D$
meson or having decayed into a muon or having interacted; 
\item[{\it cat 5:}] the $\Dz$ is not decaying semileptonically. 
\end{itemize}
Scale factors for categories 1 and 3 are fitted, 
a correction for category 4 is measured using a dedicated
event sample, whereas the factors from the two other categories are fixed to 1.0 because they contain much fewer events.    
An additional scale factor is fitted to scale the remaining
$\FourS$ background.

For the non-peaking charm background, two event categories are defined
which correspond to different distributions of the missing energy:
one for $D^0$ meson decays and the second for other charm mesons.
For the fitted values of the $p_{\pi}$-dependent scaling factors, the values of 
the two parameters for these non-peaking samples are compatible with 1.0, and they are fixed. 
The values of all those scale factors are given in Table \ref{tab:corrsb12_tight}.

\begin{table*}[!htbp]
\begin{center}
  \caption[]{Correction factors for 2-D distributions of the pion momentum and missing energy in the signal hemisphere, for sideband events: six scale factors for light-quark continuum and six for the charm non-peaking background, each for six intervals in pion momentum, five scale factors of which three are fixed for peaking charm background, two fixed scale factors for non-peaking charm background, and one scale factor is fitted for \BB\ background.  
  \label{tab:corrsb12_tight}}
\begin{tabular}{lc c c c c c c}
 \hline \hline
Background	      & 1 & 2 & 3 & 4 & 5 & 6 \\  \hline
pion momentum ($\gevc$) &     [0.0 - 0.3]& [0.3 - 0.6] & [0.6 - 0.9 ] & [0.9 - 1.2] & [1.2 - 1.5] & $\ge$ 1.5 \\ 
light quark pair bg   
               & $1.40 \pm 0.42$& $0.92 \pm 0.21$& $1.31 \pm 0.25$& $1.01 \pm 0.34$& $0.89 \pm 0.28$& $0.82 \pm 0.11$\\
charm non-peaking bg  &$1.17 \pm 0.07$& $1.02 \pm 0.05$& $0.98 \pm 0.06$& $0.78 \pm 0.07$& $0.88 \pm 0.09$& $0.76 \pm 0.07$\\
\hline
charm peaking bg&$0.94 \pm 0.13$& $1.0$ (fixed) &$0.96 \pm 0.08$&
$1.0$ (fixed) &$1.0$ (fixed) &  \\
charm non-peaking bg  &$1.01$ (fixed)& $0.97$ (fixed)& & & & \\
$ \BB$  background &$1.05 \pm 0.03$& & & & & \\
\hline \hline
\end{tabular}
\end{center}
\end{table*}

Using the error matrix from these sideband fits, the total impact of these background uncertainties is evaluated for signal event yields (S4).

As mentioned above, the corrections to the two non-peaking charm backgrounds are
 estimated from data in a first pass of the fit and then
fixed to their fitted values to obtain the $p_{\pi}$ dependent corrections. The systematic uncertainty (S5) corresponds to small changes observed when values 
of these two parameters are fixed instead to unity.

For the peaking charm-background categories 2, 4, and 5, the scale factors are fixed in the overall 2-D fit, and the assessment of the impact of fixing these scale factor is presented in the following.

For background from $\Dz \rightarrow \rho^- \ep \nue$ decays (S6), in which
the  pion originates from the $\rho$, we assess the uncertainty by 
varying the branching fraction  ${\cal B}(\Dz \rightarrow \rho^- \ep \nue)$ by
$\pm 30\%$.
This variation is larger than the present uncertainty  of $21\%$
and covers potential contributions of pions not originating from $\rho$ decays. 
Category 4 contains mainly Cabibbo-allowed decays with the charged
kaon identified as a pion. This probability is measured in data
and simulation using $\Dz \rightarrow \Km \pip$ decays and is found to be
of the order of $0.4\%$. Differences are corrected depending on the 
kaon momentum and direction measured in the laboratory. Taking into account
uncertainties in the determination of the corrections, half
of the variations on fitted quantities are used to evaluate the
corresponding systematic uncertainties (S7).

There are very few events from non-semileptonic $D^0$ decays (S8). 
Thus we choose to set the scale factor to 1.0 and assign a $30\%$ uncertainty to this source of background.

\subsubsection{Form Factors (S9-S14)}

Since semileptonic decays of $D$ and $D_s$ mesons contribute to sizable background the
knowledge of their hadronic form factors is important for the simulation of 
their $q^2$ dependence.
In Table \ref{tab:ff_syst} the  values of the relevant parameters that were recently measured by
\babar~\cite{ref:kenu,ref:dkpienu} are listed.
The simulated events were reweighted to correspond to these values.
The quoted uncertainties on these measured parameters determine the systematic uncertainties in the event yield.

\begin{table}[!htbp!]
\begin{center}
  \caption[]{ {Most recent values and uncertainties for parameters used in the simulation of the 
$q^2$ dependence  of the hadronic form factors in
semileptonic decays of $D$ and $D_s$ mesons. 
These decays are principal sources of background
as discussed in the text, and correspond to systematics S9 to S14.
$D_s \to P$ and $D \to V$  refer to decays into 
pseudoscalar and vector mesons, respectively. 
$m_P$, $m_A$ and $m_V$ correspond to the pole masses entering in the 
form factors. $A$ and $V$ refer to axial and vector form factors, respectively. 
$r_2$ and $r_V$ represent form factor ratios.  

}
  \label{tab:ff_syst}}
\begin{tabular}{l l l c}
 \hline \hline
Source & Decay & Parameters & Ref.\\
\hline
S9  & $D\rightarrow K \ep \nue$ & $ m_P = (1.884 \pm 0.019)~\gevcc $ &\cite{ref:kenu} \\
S10  & $D_s^+\rightarrow P \ep \nue$ &$m_P = (1.9 \pm 0.1)~\gevcc   $  &\\
\hline
S11  & $D\rightarrow V \ep \nue$ & $r_2 = 0.801 \pm 0.028  $ &\cite{ref:dkpienu}  \\
S12  &                            & $r_V = 1.463 \pm 0.035 $ &\\
S13  &                            & $m_A = (2.63 \pm 0.16)~\gevcc   $ &\\
S14  &                            & $m_V = (2.1  \pm 0.2)~\gevcc   $ &\\
\hline \hline
\end{tabular}
\end{center} 
\end{table}

\subsubsection{$\Dz$ Reconstruction (S15)}
The measurement of the $\Dz$ direction and energy is critical 
for the $q^2$ determination. The reference sample
of $\Dz \to \Km \pip$ decays has shown rather small differences between 
data and simulation and these have been corrected in the simulation
of the signal and reference samples~\cite{ref:kenu,ref:dskkenu,ref:dkpienu}. 
We adopt as systematic uncertainties the changes in the results
obtained with and without these corrections.
 
\subsubsection{Electron Identification (S16)}

Differences between data and simulated events for the electron
identification are corrected using \babar\ standard procedures.
The impact of these corrections is taken as an estimate of 
the systematic uncertainty.

\subsubsection{Radiative Corrections (S17)}

Effects of initial and final state radiation are simulated using PHOTOS
\cite{ref:photos}.
By comparing two generators (PHOTOS and KLOR \cite{ref:klor}), the CLEO-c collaboration has used a variation of $16\%$ to evaluate the corresponding systematic uncertainty \cite{ref:cleorad}. We have changed the fraction of radiative events by $30\%$ (keeping constant the total number of events) and obtained the corresponding variations on fitted parameters.

\subsubsection{Pion Identification (S18)}

Stringent requirements on pion identification are applied to reduce background from the Cabibbo-favored
$\Dz \to \Km \ep \nue$ decays. The efficiency of the particle identification (PID) algorithm as a function of the pion momentum and polar angle in the laboratory frame is studied on the data and MC samples for
$\Dstarp \rightarrow \Dz \pi_s^+,~ \Dz \rightarrow \Km \pip$ decays.
Specifically, the pion from the $\Dz$ decays is selected without any
PID requirement, as
the track with the same charge as the $\pi_s^+$ from the $\Dstarp$ decay.

\begin{figure}[!htbp!]
\centering
\includegraphics[width=9cm]{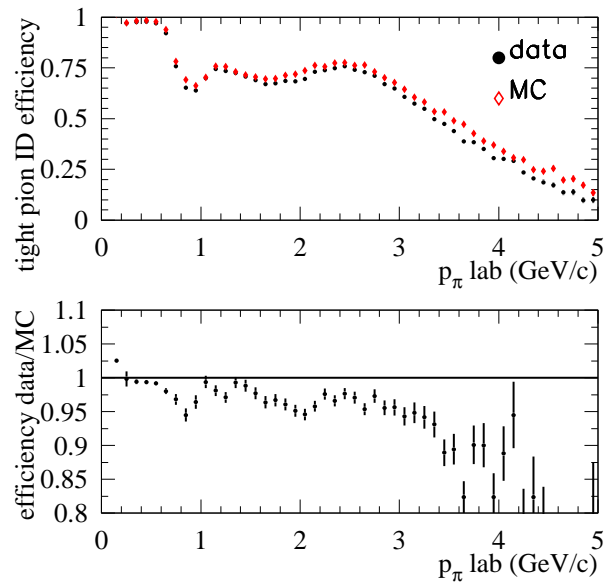}
\caption{{ Study of the uncorrected pion efficiency in data and in simulation
versus the laboratory pion momentum; top: measured efficiencies, bottom: ratio of efficiencies in data and MC.
}
\label{fig:tight_eff_plab}}
\end{figure}

For data and MC-simulated events, Fig.~\ref{fig:tight_eff_plab} shows a comparison of the measured pion efficiency as a function of the pion momentum in the
laboratory.
After applying corrections, which depend on the track momentum and
angle measured in the laboratory, these differences are reduced by a factor 
five. The systematic uncertainty related to these corrections is obtained
by scaling the variations on measured quantities, before and after 
corrections, by this same amount.

\subsubsection{$q^2$ Reconstruction (S19)}

As part of the previous \babar\ analysis of $\Dz \rightarrow \Km \ep \nue$ \cite{ref:kenu} decays,
we studied the variation of the efficiency versus $q^2$ in data and simulation.
For this purpose, $\Dz \rightarrow \Km \pip \piz$ decays were analyzed, ignoring the $\piz$, but otherwise using the standard algorithm for semileptonic $\Dz$ decays.
No significant difference was observed and a straight-line was fitted
to the ratio of the efficiency in data and simulation.
To assess the systematic uncertainty on the current measurement related to 
this effect, we vary the slope of the $q^2$ distribution by $1\%$, leaving the total number of selected events unchanged. No correction is applied
to the $q^2$ variation because the measured effect is compatible with its uncertainty.

\subsubsection{Kaon Veto (S20)}

The relatively tight PID requirement for the signal charged pion is combined with a loose kaon selection to veto $\Km$.  Specifically, among events with at least one charged particle in the candidate hemisphere, 
in addition to the $\pim$ and $\pi^+_s$, the particle is assumed to be a kaon if  it is oppositely charged relative to the $\pi_s^+$ from the $\Dstarp$, has a momentum of at least $400 \mevc$, and passes loose requirements for kaon identification.  Such events are vetoed. Based on the same method employed for charged pions, we confirm very good data-MC agreement.
For example, 
the ratio of efficiencies measured
in data and simulation is equal to $1.005\pm0.001$.
A small difference measured for kaons of momentum smaller than $800 \,\mevc $ 
is corrected. 
The systematic uncertainty corresponding to the changes in the veto efficiency  for low-momentum, loosely identified kaons is adopted.

\subsubsection{Cross Check}

\begin{figure*}[!htb!]
  \begin{center}
    \mbox{\epsfig{file=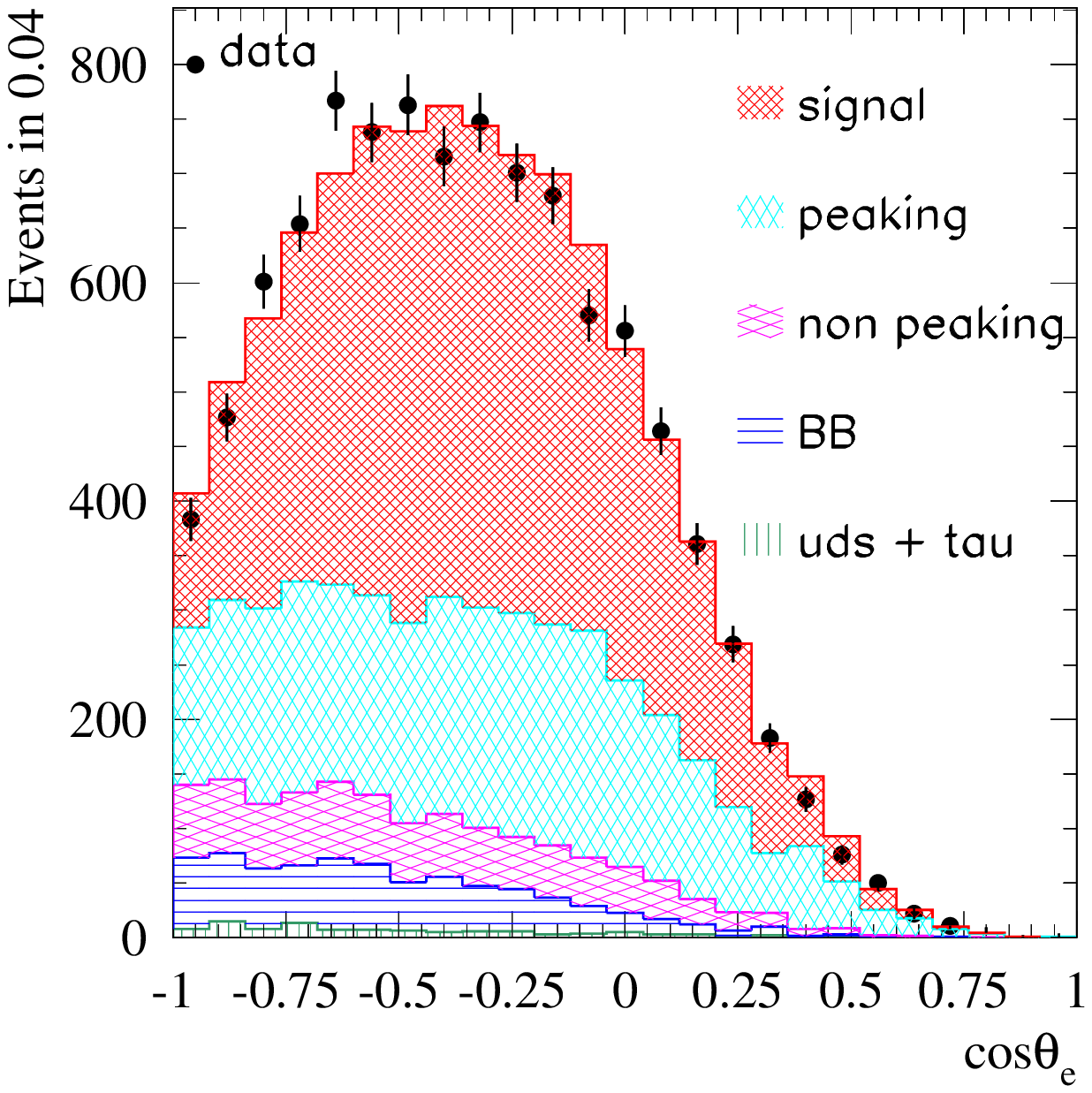,width=0.5\textwidth}
\epsfig{file=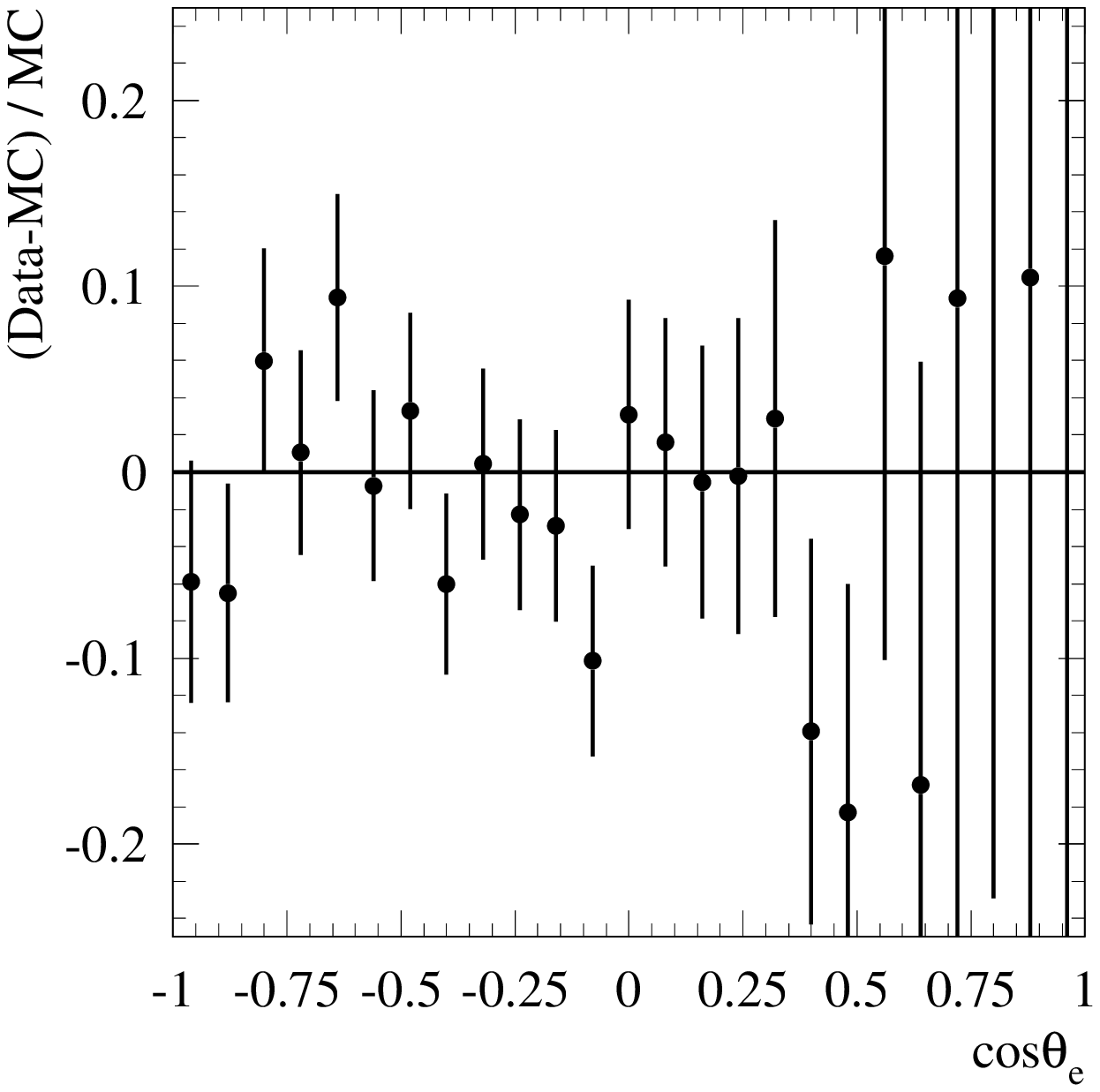,width=0.5\textwidth}}
  \end{center}
\caption{{ Comparison of the measured event yields  (black data points with statistical errors),  as a function of
$\cos{\theta_e}$, with the corrected sum of the expected signal and background distributions after all corrections. Left: observed events in data and in simulation. Right: the ratio (Data - MC)/MC. 
\label{fig:fitted_cthe_tight}}}
\end{figure*}

The distribution of the helicity angle, $\theta_e$, 
is determined by the dynamics of the $V-A$ interaction for a decay to a pseudoscalar meson. 
Figure\,\ref{fig:fitted_cthe_tight} shows a comparison of the selected event yields and the sum of the expected signal and background contributions as a function of  $\cos{\theta_e}$. 
As in Figure \ref{fig:q2rall}, this distribution is not corrected for efficiency and resolution effects.  
The helicity angle $\theta_e$ is not used to evaluate any of the corrections
to the simulation. Therefore, this figure illustrates independently the very good agreement between data and the corrected simulation. 
Furthermore, the ratio (Data - MC)/MC shows no significant dependence on 
$\cos{\theta_e}$;  a fit to a constant results in 
$(-1.5 \pm 1.3)\times 10^{-2}$ and a $\chi^2/\ndf = 18.8/24$.

\section{Results}

So far, we have presented the observed $q^2$ (see Fig.~\ref{fig:q2rall}) and helicity distributions (see Fig.~\ref{fig:fitted_cthe_tight}).  
The background-subtracted $q^2$ distribution is unfolded to take into account the detection efficiency and resolution effects 
(see Table~\ref{tab:q2_unfolded_tight}).   The systematic uncertainties on the unfolded yields are evaluated in ten discrete intervals of $q^2$ (see Table~\ref{tab:syst1_tight_unfold}).  In the following, we discuss the measurements of the integrated branching fraction, the $q^2$ distribution, and the measurement of the hadronic form factor.

\subsection{Branching Fraction Measurement}
\label{sec:rate}

As the primary result of this analysis  we present the ratio of branching fractions,
\begin{equation}
R_D = \frac { {\cal B}(\Dz \to \pim \ep \nue)} {{\cal B}(\Dz \to \Km \pip)},
\end{equation}
i.e., the signal semileptonic decay $\Dz \to \pim \ep \nue$ measured
relative to the hadronic decay $\Dz \to \Km \pip$.
In both channels the $\Dz$ originates from a $\Dstarp$ decay and photons 
radiated in the final state are taken into account.
The same ratio $R_D$ is obtained for simulated event samples, many systematic uncertainties cancel. The signal decay branching fraction
is obtained by multiplying $R_D$
by the branching fraction for the hadronic decay $\Dz \rightarrow \Km \pip$~\cite{ref:hfag12},
\begin{equation}
{\scriptsize
{\cal B}(\Dz \to \Km \pip)_{{\rm WA}}= (3.946 \pm 0.023 \pm 0.040 \pm 0.025 ) \% ,}
\label{eq:bkpidata}
\end{equation}
\noindent
where the stated first uncertainty is statistical, the second systematic, and last includes the effect of modeling final state radiation.
The measurement of the ratio $R_D$ is detailed in the following way,
\begin{eqnarray}
R_D &=& \frac{{\cal B}(\Dz \rightarrow \pim \ep \nue)_{{\rm data}}}{{\cal B}(\Dz \rightarrow \Km \pip)_{{\rm WA}}}\label{eq:rd}\\
&=&N(\pim \ep \nue)_{{\rm data}}^{{\rm corr.}} \frac{N(\Km \pip)_{{\rm MC}}}{N(\Km \pip)_{{\rm data}}}
\frac{{\cal L}({\rm data})_{K\pi}}{{\cal L}({\rm data})_{\pi e\nu}}\nonumber\\
&\times&\frac{1}{2\,{N}({\rm MC})_{K\pi}}\,R_{\epsilon} \frac{1}{\epsilon_{{\rm had.}}}\,\frac{1}{{\cal P}(c \rightarrow \Dstarp)_{{\rm MC}}} \nonumber\\
&\times&\frac{1}{{\cal B}(\Dstarp \rightarrow \Dz \pip)_{{\rm MC}}\,{\cal B}(\Dz \rightarrow \Km \pip)_{{\rm MC}}}.\nonumber
\end{eqnarray}

In this expression,
\begin{itemize}
\item $N(\pim \ep \nue)_{{\rm data}}^{{\rm corr.}}=\frac{N(\pim \ep\nue)_{{\rm data}}}
{\epsilon(\pim \ep\nue)_{{\rm MC}}}$ is the number of unfolded signal events
(see Table~\ref{tab:q2_unfolded_tight});

\item $N(\Km\pip)_{{\rm MC}}$ and $N(\Km \pip)_{{\rm data}}$ are
the numbers of measured events in simulation and data, respectively;

\item ${\cal L}({\rm data})_{\pi e\nu}=347.2~\fb^{-1}$ and
${\cal L}({\rm data})_{K\pi}=92.89~\fb^{-1}$ refer to the integrated luminosities analyzed for the signal
and the reference decay channels, respectively;

\item ${N}({\rm MC})_{K\pi}=152.4 \times 10^6$ refers to the total number of $e^+e^- \to \ccbar$ simulated events
with a $\Dz \to \Km \pip$ decay;
\item $R_{\epsilon}$ is the double ratio of efficiencies to reconstruct
signal events in the two decay channels in data and simulation;

\item $\epsilon_{{\rm had.}}=0.9596$ is the hadronic tagging efficiency
which is included in the simulation for the reference channel, but not for the signal channel.

\item ${\cal P}(\c \rightarrow \Dstarp)_{{\rm MC}}=0.2307$ is the probability
for a $\c$ quark to produce a $\Dstarp$ meson;

\item ${\cal B}(\Dstarp \rightarrow \Dz \pip)_{{\rm MC}}=0.683$ is the branching
fraction assumed in the MC;

\item ${\cal B}(\Dz \rightarrow \Km \pip)_{{\rm MC}}=0.0383$ is the branching
fraction assumed in the MC.

\end{itemize}

\begin{figure}[!htb]
  \begin{center}
\includegraphics[width=9cm]{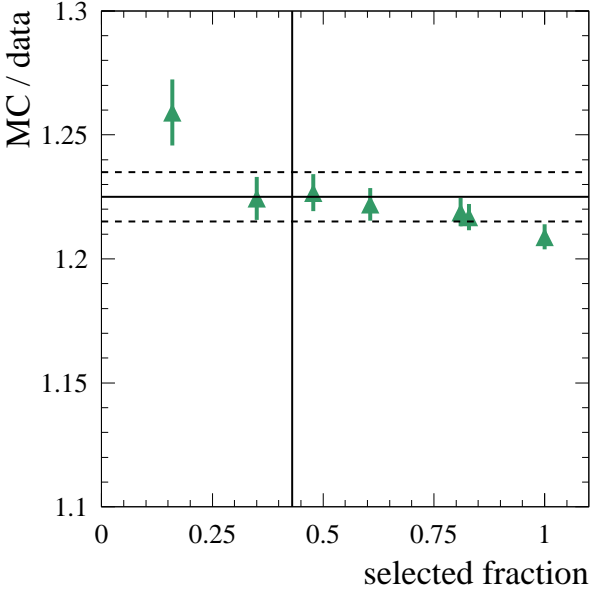}
  \end{center}
  \caption[]{ Variation of the ratio of numbers of $\Dz \to \Km \pip$ 
events measured in MC and data, as a fraction of events selected
by restrictions on $F_{\ccbar -2}$. The vertical
line shows the fraction of selected $\Dz \rightarrow \pim \ep \nue$
events after the requirement on $F_{\ccbar}$. All events satisfy $F_{\bbbar}>1.2$.
The horizontal lines indicate the value adopted for this ratio (full line)
and the corresponding uncertainty (dashed lines).}
  \label{fig:stab_ratio_fcc}
\end{figure}

To cancel a large fraction of systematic uncertainties,
similar selection criteria are used for the two $\Dz$ decays.
The following criteria are common for the selection of the
two channels:

\begin{itemize}
\item {\it Particle identification.}  
The pion identification of both decay channels is the same, and
no identification is requested for the kaon in the $\Dz \to \Km \pip$ decay.

\item {\it Global event topology.} The event selection for the two decay channels are analyzed in the same way. Specifically, we only retain events with $|\cos{\theta_{{\rm thrust}}}|<0.6$ and a missing energy in the opposite hemisphere of less than 3 $\gev$.

\item {\it Fragmentation-related variables.} For the two channels, we 
require at least one spectator particle in the signal candidate hemisphere and apply the same veto against additional kaons in that
hemisphere. 

\item {\it Vertexing.} For the probability of the 
$\Dz$ and $\Dstarp$ decay vertex fits,  we require $P(\chi^2) > 0.01$.
We also discard events with the distance of closest
approach in the transverse plane that exceeds 1\mm ,
for the pion trajectory relative to the interaction vertex.

\item {\it Fisher variables.} The same restriction on the Fisher discriminant
$F_{\bbbar}$ is used to suppress \BB\ background. For continuum suppression 
in the hadronic $\Dz$ decay sample, we replace the eight-variable Fisher discriminant $F_{\ccbar}$ with the six-variable discriminant 
$F_{\ccbar -2}$, which does not include the two variables related to the final state electron.
We have verified the stability of the result with respect to 
a restriction  on $F_{\ccbar -2}$,  as shown 
in Fig.\,\ref{fig:stab_ratio_fcc}.
The value 
of $N(\Km \pip)_{{\rm MC}}/N(\Km \pip)_{{\rm data}} = 1.225 \pm 0.008 \pm 0.010$ covers the variation of this ratio for a wide range of restrictions on $F_{\ccbar -2}$.
\end{itemize}

\noindent

The ratio of efficiencies,
\begin{equation}
R_{\epsilon} = \frac{\epsilon(\Km \pip)_{{\rm data}}}{\epsilon(\Km \pip)_{{\rm MC}}}
\frac{\epsilon(\pim \ep \nue)_{{\rm MC}}}{\epsilon(\pim \ep \nue)_{{\rm data}}}
=1.006 \pm 0.007,
\end{equation}
\noindent
is only impacted by event selection criteria that are different for the two decay channels, specifically  
\begin{itemize} 
\item limits on the $\Km \pip$ invariant mass and on the mass difference
$m_{K^-\pi^+\pi^+_s}-m_{K^-\pi^+}$;
\item limits on the mass difference $\Delta(m)= m_{D^0\pi^+_s}-m_{D^0}$ 
after the first kinematic fit (see Fig.\,\ref{fig:deltam});
\item limits on $\chi^2$ probabilities for the two kinematic fits. 
\end{itemize}

The impact of differences between data and simulated events has been
assessed based on the earlier $\DdecK$ measurement~\cite{ref:kenu}.

Common sources of systematic uncertainties (S2, S18, and S20)
contributing to the measured number of unfolded signal events
$(N(\pim \ep \nue)_{{\rm data}}^{{\rm corr.}})$ and  the ratio of reconstructed
$\Km \pip$ events ($N(\Km \pip)_{{\rm MC}}/N(\Km \pip)_{{\rm data}}$)
are evaluated taking into account correlations. 

Based on the total number of efficiency corrected signal events,
\begin{equation}
N(\pim \ep \nue)_{{\rm data}}^{{\rm corr.}}=(375.4 \pm 9.2 \pm 10.1) \times 10^3 ,
\end{equation} 
we obtain for the ratio of branching fractions,
\begin{equation}
 R_D  = 0.0702 \pm 0.0017 \pm 0.0023, 
\end{equation}
where the first uncertainty is statistical and the second is systematic. 
Using the $\Dz \rightarrow \Km \pip$ branching fraction, given in 
Eq.~(\ref{eq:bkpidata}), we arrive at
\begin{equation}
{\cal B}(\Dz \rightarrow \pim \ep \nue)= (2.770 \pm 0.068 \pm 0.092 \pm 0.037)\times 10^{-3},  
\end{equation}
where the third error accounts for the uncertainty 
on the branching fraction for the reference channel.
This value is slightly lower, but consistent with the present world average of
$(2.89 \pm 0.08)\times 10^{-3}$ \cite{ref:pdg2013}.

\begin{table*}[htbp]
  \caption {Differential branching fractions $(\Delta{\cal B}(\Ddecpi))$ 
in ten bins in $q^2$, spanning from 0 to $q^2_{\rm max}$ in $\gev^2$ (second row), with separate statistical and systematic uncertainties and correlation matrices below.  
The second row lists the values of the differential branching fraction 
integrated over 0.3 $\gev^2$ intervals (quoted in the first row). 
The off-diagonal elements of the correlation matrices are provided for both the statistical (upper half) and systematic (lower half) uncertainties. 
The diagonal elements refer to the  uncertainties ($\times 10^3$).
The uncertainty on the normalization channel (see Eq. (\ref{eq:bkpidata})) 
must be added when evaluating the total uncertainty.
}
\begin{center}
\small{
  \begin{tabular}{lcccccccccc}
    \hline\hline
$q^2$ bin $(\gev^2)$& [0.0, 0.3] &[0.3, 0.6] &[0.6, 0.9] &[0.9, 1.2] &[1.2, 1.5] &[1.5, 1.8] &[1.8, 2.1] &[2.1, 2.4] &[2.4, 2.7] &[2.7, $q^2_{\rm max}$] \\
\hline
$\Delta{\cal B}\times 10^3$ &0.5037 &0.4672  & 0.4551&  0.3827  & 0.3037&0.2664  &0.2110 &0.1235  &0.0477 & 0.0090  \\ \\
\hline
stat. &0.0257 &    -0.3345 &    -0.1429 &     0.0732 &     0.0121 & -0.0097 &    -0.0024 &     0.0004 &     0.0004 &     0.0003 \\
uncert. &  &     0.0315 &    -0.1420 &  -0.2417 & 0.0401 & 0.0311 & -0.0034 &    -0.0050 &    -0.0007 &     0.0003   \\
and  &  &  &     0.0290 &    -0.0852 &    -0.2376 & 0.0205 & 0.0368 &0.0034 &    -0.0062 & -0.0062  \\
correl. &  &  &   &     0.0283 & -0.0110 & -0.2395 & -0.0223 & 0.0330& 0.0119 &     0.0034 \\
  &  & &  &   & 0.0263 &  0.0702 & -0.2221 & -0.0600 &0.0281 & 0.0382  \\
 &  & &  &   &  & 0.0254 & 0.2619 & -0.1551 & -0.1050 & -0.0614  \\
 &  & &  &   &  &  &     0.0239 & 0.3904 & -0.1211 & -0.2012  \\
 &  & &  &   &  &  &  &  0.0200 & 0.5148 &  0.2643   \\
 &  & &  &   &  &  &  &  &  0.0174 & 0.9233   \\
 &  & &  &   &  &  &  &  &  &     0.0057 \\   \\
\hline
syst. &     0.0133 &     0.7488 &     0.7239 &     0.6568 &     0.6321 &      0.3769 &     0.0735 &     0.0309 &     0.1667 &     0.2194 \\ 
uncert. &  & 0.0174 &     0.8281 & 0.3433 & 0.6907 & 0.4597 & 0.1576 & 0.1800 &     0.3585 & 0.4216 \\
 and &  &  & 0.0136 & 0.4608 & 0.6949 & 0.6524 & 0.3740 & 0.3482 & 0.4333 & 0.4196\\
correl. & & & & 0.0119 & 0.7096 & 0.4462 & 0.2939 & 0.2055 & 0.2310 & 0.1772 \\
& & & & & 0.0103 & 0.7076 & 0.4513 & 0.4597 & 0.6588 & 0.6371 \\
& & & & & & 0.0120 & 0.8772 & 0.8344 & 0.7076 & 0.5088 \\
& & & & & & & 0.0181 & 0.9644 & 0.6184 & 0.3135 \\
& & & & & & & & 0.0156 & 0.7439 & 0.4539 \\
& & & & & & & & & 0.0087 & 0.9345 \\
& & & & & & & & & & 0.0025 \\ \\
    \hline\hline
  \end{tabular}}
\end{center}
  \label{tab:errmeas}
\end{table*}

\subsection{Differential Decay Rate and Normalization}
\label{sec:diff_rate}

\begin{figure}[!htb]
\begin{center}
\includegraphics[height=9cm]{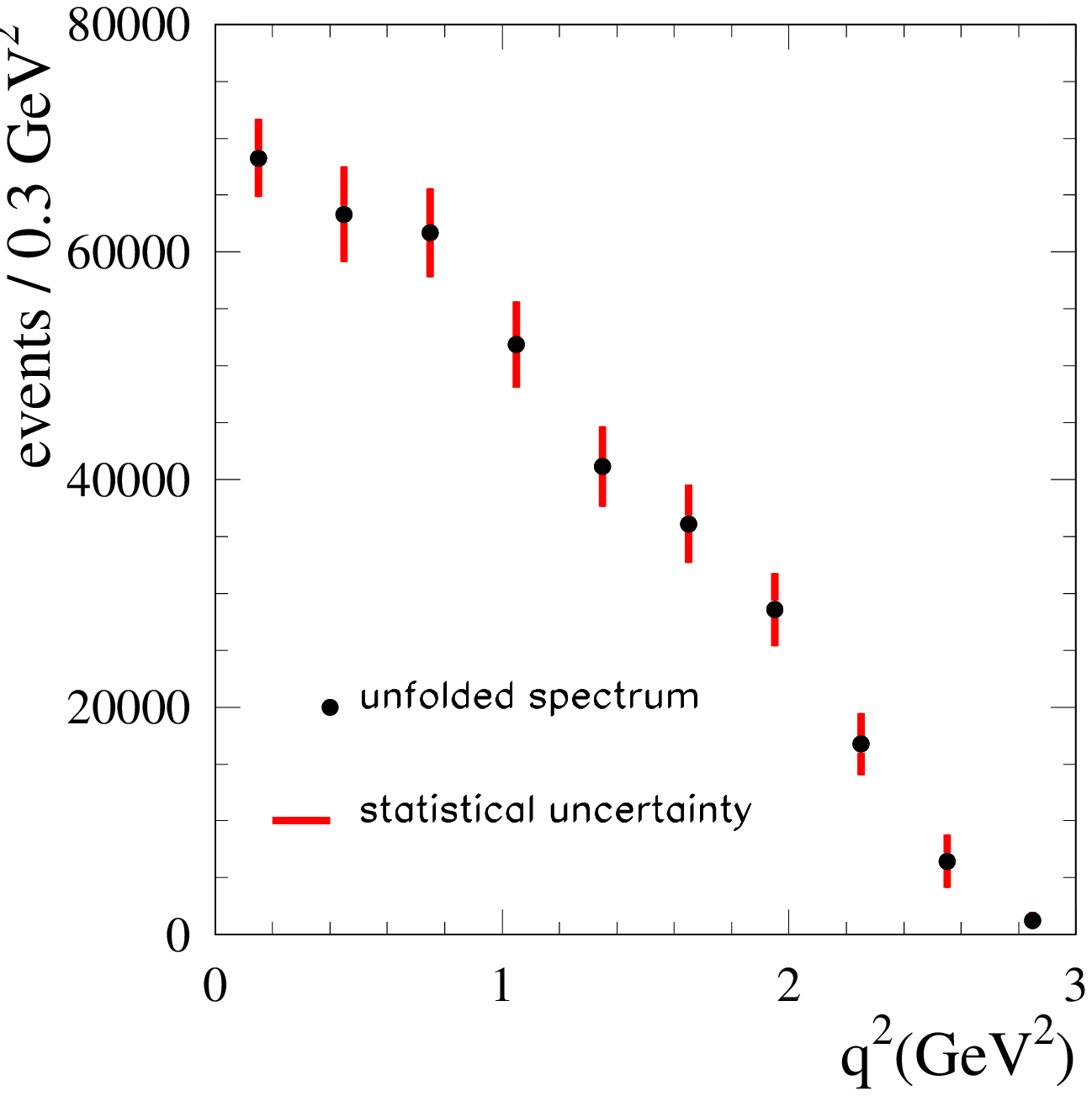}
\caption{ Unfolded $q^2$ distribution for $\Dz \to \ep \pim \nue$ decays.
}
\label{fig:q2datasim}
\end{center}
\end{figure}

 Figure~\ref{fig:q2datasim} shows the background-subtracted unfolded $q^2$ distribution. The unfolding takes into account detection efficiency correction and resolution effects.   
Based on the unfolded $q^2$ distribution  and the detailed analysis of the systematic uncertainties as a function of $q^2$ presented in 
Table~\ref{tab:syst1_tight_unfold},  Table \,\ref{tab:errmeas} lists the 
partial differential branching fractions  $\Delta{\cal B}(\Ddecpi)$ in ten $q^2$ intervals, together with the   
statistical and systematic uncertainties and the correlation coefficients.
Correlations between systematic uncertainties for neighboring $q^2$ intervals
are sizable. 
Note that the partial decay branching fractions in each $q^2$ interval
are corrected for radiative effects and that the uncertainty on the 
normalization channel (see Eq. (\ref{eq:bkpidata})), which is common to all 
ten measurements, is not included
in the uncertainties in Table \,\ref{tab:errmeas}.

The overall decay rate is proportional to the square of the product
$\Vcd \times f_{+,D}^{\pi}(q^2)$, with the $q^2$ dependence determined by the 
form factor. Its value at $q^2=0$ can be expressed as,
\begin{equation}
\Vcd \times f_{+,D}^{\pi}(0)=
\sqrt{\frac{24 \pi^3}{G_F^2}\frac{{\cal B}(\Ddecpi)}{\tau_{D^0} \, I}},
\end{equation}
\noindent where
$\tau_{D^0}=(410.1\pm1.5)\times10^{-15}~s$ \cite{ref:pdg2013} is the $\Dz$ lifetime and
$I=\int_0^{q^2_{{\rm max}}}{\left | \vec{p^*_{\pi}} (q^2) \right |^3 \left |f_{+,D}^{\pi}(q^2)/f_{+,D}^{\pi}(0) \right |^2}~dq^2$.
Based on the $z$-expansion parameterization of the form factor, we determine the integral $I$ and obtain,
\begin{equation}
\Vcd \times f_{+,D}^{\pi}(0)= 0.1374 \pm 0.0038 \pm 0.0022 \pm 0.0009,
\label{eq:vcdf0}
\end{equation}
where the third uncertainty corresponds to the uncertainties
on the branching fraction of the normalization channel $\Dz \to K^- \pi^+$  and
on the $\Dz$ lifetime.

From the measured branching fraction (Table \,\ref{tab:errmeas}) as a function of
$q^2$ intervals,  $\Vcd \times f_{+,D}^{\pi}(q^2)$ is derived and shown in Fig.\,\ref{fig:babar_taylor},
where the data are evaluated at the center of each $q^2$ bin (see Appendix \ref{sec:appendixd}). 
The data are compared to the fit  based on the $z$-expansion parameterization of the form factor
with three free parameters, the normalization $\Vcd \times f_{+,D}^{\pi}(q^2=0)$  and the shape parameters $r_1$ and $r_2$. They are 
considered in that order in the following. The correlation coefficients $(\rho_{ij})$ are $\rho_{12}=-0.400$, $\rho_{13}=0.572$, and $\rho_{23}=-0.966$. The form factor fit reproduces the data well, $\chi^2=2.6$ for 7 degrees of freedom.

\begin{figure}[!htb]
  \begin{center}
\includegraphics[width=9cm]{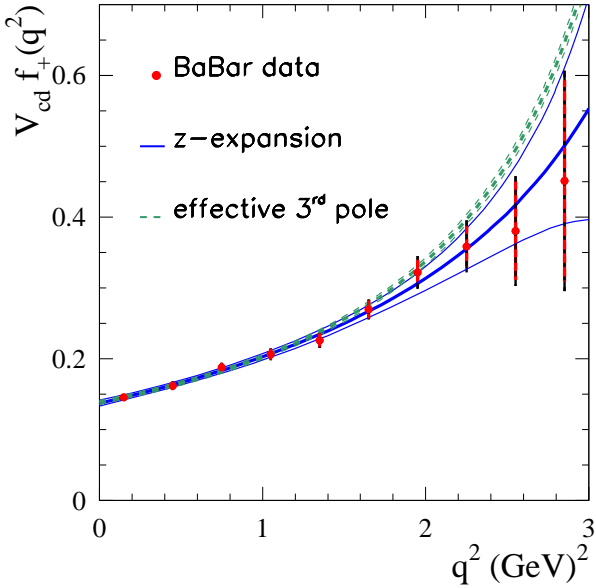}
  \end{center}
  \caption[]{Measured values of $\Vcd \times f_{+,D}^{\pi}(q^2)$
are compared with the results of a fit using a $z$-expansion parameterization of
the hadronic form factor (full blue line). The dashed (green)
lines show the comparison with a fit using 
the effective three-pole ansatz in which the mass of an effective third pole is fitted. The superconvergence condition and constraints on the two 
first residues are imposed (see Section\ref{sec:threepoles_physics}).}
  \label{fig:babar_taylor}
\end{figure}

Using a recent unquenched lattice LQCD computations of
the hadronic form factor, $f_{+,D}^{\pi}(0)= 0.666\pm 0.029$ \cite{ref:hpqcd1},
we obtain a value for the CKM matrix element, 
\begin{equation}
\Vcd = 0.206 \pm 0.007_{{\rm exp.}} \pm 0.009_{{\rm LQCD}} ,
\end{equation}
where the first uncertainty is the quadratic sum of the statistical and 
systematic measurement uncertainties, and the second corresponds to 
uncertainties on the LQCD prediction.

If, instead, we use $\Vcd=\Vus=\lambda$, 
the normalization of the hadronic form factor becomes, 
\begin{equation}
f_{+,D}^{\pi}(0)= 0.610 \pm 0.020_{{\rm exp.}} \pm 0.005_{{\rm other}},
\end{equation}
where the first uncertainty corresponds to statistical and systematic
uncertainties given in Eq. \,(\ref{eq:vcdf0}). The second uncertainty 
corresponds to the uncertainties
on the branching fraction of the normalization channel,
on the $\Dz$ lifetime, and on $\Vcd$.

The measurements presented here are compared in Table \,\ref{tab:normresults} with previous results from other experiments which were also
based on the three-parameter fit of the $z$-expansion
parameterization of the hadronic form factor.  The results are consistent within the stated uncertainties.  The sizable variation of the fitted shape parameters $r_1$ and $r_2$ can be traced to the large experimental uncertainties at high $q^2$, the correlation is almost $100\%$ between these two quantities. In the comparison with LQCD estimates, the value of $\Vcd=\Vus=0.2252\pm0.0009$ 
is used.  

\begin{table*}[!htb]
\begin{center}
 \caption[]{Measurements of the normalization factor
$\Vcd \times f_{+,D}^{\pi}(0)$ and of the parameters $r_1$ and $r_2$ 
used in the $z$-expansion parameterization of the 
hadronic form factor. The two sets of values for the CLEO-c (2008) untagged analysis
correspond to the $\pim \ep \nue$ and $\piz \ep \nue$
channels, respectively. 
Predictions based on four LQCD calculations, obtained using $\Vcd=\Vus$, are listed at the bottom.}
  \label{tab:normresults}
\begin{tabular}{lc c c c c}
\hline\hline
Experiment & ref. & $\Vcd \times f_{+,D}^{\pi}(0)$& $r_1$ &$r_2$ \\
\hline 
Belle (2006)&\cite{ref:belle} & $0.140 \pm 0.004 \pm 0.007 $ & &\\
CLEO-c untagged (2008) &\cite{ref:cleoc08} &  $0.140 \pm 0.007\pm 0.003$ &$-2.1\pm0.7$ &$-1.2\pm4.8$ \\
CLEO-c untagged (2008) &\cite{ref:cleoc08} &  $0.138 \pm 0.011\pm 0.004$ &$-0.22\pm1.51$ &$-9.8\pm9.1$ \\
CLEO-c tagged (2009) &\cite{ref:cleoc09} &  $0.150 \pm 0.004\pm 0.001$ &$-2.35\pm0.43\pm0.07$ &$3\pm3$ \\
BESIII (2012)(prel.) &\cite{ref:bes312} & $0.144 \pm 0.005\pm 0.002$ &$-2.73\pm0.48\pm0.08$ & $4.2\pm3.1 \pm0.4$ \\
\hline
HFAG average (2012) &\cite{ref:hfag12} & $0.146 \pm 0.003$ & $-2.69 \pm 0.32$ & $4.18 \pm 2.16 $\\
\hline
BESIII (2014)(prel.) &\cite{ref:bes3} & $0.1420 \pm 0.0024\pm 0.0010$ &$-1.84\pm0.22\pm0.07$ & $-1.4\pm1.5 \pm0.5$ \\
\hline
This analysis & & $0.137 \pm 0.004 \pm0.002 \pm0.001$ &$-1.31\pm0.70\pm0.43$ &$-4.2\pm4.0\pm1.9$ \\
\hline\hline
  & & & & \\  \hline\hline
LQCD Predictions &  ref. & $\Vcd \times f_{+,D}^{\pi}(0)$& $r_1$ &$r_2$ \\ 
\hline  
FNAL/MILC (2004) &\cite{ref:fnal_milc} & $0.144 \pm 0.016$ & &\\
ETMC (2011)&\cite{ref:etmc1} & $0.146 \pm 0.020$ & &\\
HPQCD (2011)&\cite{ref:hpqcd1} & $0.150 \pm 0.007$ & &\\
HPQCD (2013)&\cite{ref:hpqcd2} & $0.153 \pm 0.009$ & $-1.93\pm0.20$&
$0.37 \pm 0.93$\\
\hline\hline
\end{tabular}
\end{center}
\end{table*}

\begin{figure}[!htb]
  \begin{center}
\includegraphics[width=9cm]{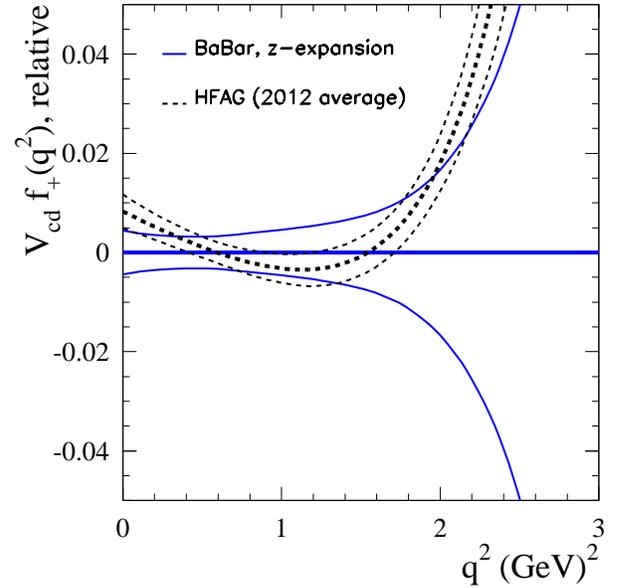}
  \end{center}
  \caption[]{ Comparison of this measurement with an average by HFAG of
all other results listed in Table~\ref{tab:normresults}, both obtained from 
fits to the $z$-expansion.
For the curves and their error bands the \babar\ results in Fig. \ref{fig:babar_taylor} 
have been subtracted.  The continuous (blue) lines illustrate total uncertainties 
on the \babar\ measurement, 
for which the central values are, by construction, equal to zero.
Dashed (black) lines are the results of HFAG.}
\label{fig:babar_others}
\end{figure}

Figure\,\ref{fig:babar_others} shows two fits to
 $\Vcd \times f_{+,D}^{\pi}(q^2)$
 based on the $z$-expansion, one for this analysis, the other
for the HFAG averaged measurements~\cite{ref:hfag12}, both
listed in Table~\ref{tab:normresults}.
 
To extract the value of $\Vcd$ we rely on a prediction from lattice QCD, 
which is the only approach to compute
$f_{+,D}^{\pi}(q^2)$ and $f_{0,D}^{\pi}(q^2)$ from first principles. 
Values of the hadronic form factor at $q^2=0$ are derived with the constraint
$f_{+,D}^{\pi}(0)=f_{0,D}^{\pi}(0)$. Recent results are listed in
Table~\ref{tab:normresults}, obtained assuming $\Vcd=\Vus=0.2252 \pm 0.0009$. 
For the evaluation of the $q^2$ dependence of the form factor  
we rely on the preliminary results from the HPQCD 
Collaboration \cite{ref:hpqcd2}.

The most precise unquenched LQCD calculations
by the HPQCD collaboration is $f_{+,D}^{\pi}(q^2=0)=0.666 \pm 0.029$
~\cite{ref:hpqcd1}.
Using this value we obtain a value for the CKM matrix element, 
\begin{equation}
\Vcd = 0.206 \pm 0.007_{{\rm exp.}} \pm 0.009_{{\rm LQCD}},
\end{equation}
where the first uncertainty corresponds to uncertainties on this measurement,
summed in quadrature, and the second to the uncertainty of the LQCD prediction.

If, instead, we adopt the value of $\Vcd=\Vus=\lambda$, 
the normalization of the hadronic form factor becomes,
\begin{equation}
f_{+,D}^{\pi}(0)= 0.610 \pm 0.020_{{\rm exp.}} \pm 0.005_{{\rm other}}.
\end{equation}
The second uncertainty corresponds to the uncertainties on the branching fraction of the normalization channel, on the $\Dz$ lifetime, and on $\Vcd$.

\subsection{Parameterization of the form factor $f_{+,D}^{\pi}(q^2)$}
\label{sec:ffq2}

\subsubsection{Fits to the $q^2$ dependence of $f_{+,D}^{\pi}(q^2)$}
\label{sec:ff0}

A summary of the fits to the $q^2$ dependence of $f_{+,D}^{\pi}(q^2)$, based
on different parameterizations is given in Table \ref{tab:fittedparam}. 
Overall, the fits describe the data well.  

Figure~\ref{fig:babar_taylor} compares the result of the fit to the $z$-expansion with a fit 
to the data based on the effective three-pole ansatz with superconvergence constraints.
Below $2 \gev^2$ the two fits agree well, at higher $q^2$ the pole fit 
lies about one standard deviation above the data, similar to the HFAG fit shown in Fig.~\ref{fig:babar_others}.

\begin{table*}
  \caption {Fitted values of the parameters corresponding to 
different parameterizations of $f_{+,D}^{\pi}(q^2)$. The last column
gives expected values for the parameters when available.}
\begin{center}
  \begin{tabular}{llll}
    \hline\hline
Ansatz  & Fitted Parameters&$\chi^2$/\ndf& Predictions \\

\hline
$z$-expansion&  $r_1  = -1.31 \pm 0.70 \pm 0.43$ &2.0/7 &  \\
   &                     $r_2  = -4.2 \pm 4.0 \pm 1.9$& &\\
\\
effective three-pole &  $m_{\rm pole3}=(3.55\pm0.30\pm0.05)\,\gevcc$ & 4.8/9&  $m_{\rm pole3}>3.1\,\gevcc$\\
\\
fixed three-pole &  $c_2=0.17 \pm 0.06 \pm 0.01$ & 3.3/7& \\
          &   $c_3=0.15 \pm 0.09 \pm 0.06$ & & \\
\\
two-pole & $b_2=1.643 \pm 0.060 \pm 0.035$ & 3.7/7& \\
          &  $b_3=0.68 \pm 0.13 \pm 0.11$ & &$~0.6$ \\
\\
modified-pole &$\alpha_{\rm pole} = 0.268 \pm 0.074 \pm 0.059$ & 3.0/8& $<0.6$\\
\\
single-pole &$m_{\rm pole} = (1.906 \pm 0.029 \pm 0.023)\,\gevcc$&5.5/8& $2.010\,\gevcc$\\
\\
ISGW2 & $\alpha_I = (0.339 \pm 0.029 \pm 0.025)\,\gev^{-2}$ &2.1/8 &$0.104\,\gev^{-2}$\\
\hline \hline
  \end{tabular}
\end{center}
  \label{tab:fittedparam}
\end{table*}

\subsubsection{Evidence for three or more poles contributions to $f_{+,D}^{\pi}(q^2)$}
\label{sec:ffq2_model}

As was pointed out in Section \ref{sec:burdman}, 
the contributions from the first two poles entering the expression
for $f_{+,D}^{\pi}(q^2)$ can be estimated using the measured masses and widths of the
$\Dstarp$ and $D^{*\prime}_1$ resonances. 
By comparison with data, these estimates can be validated and the different hadronic states
which contribute to the hadronic form factor can be identified.

\begin{figure}[!htb]
  \begin{center}
\includegraphics[width=9cm]{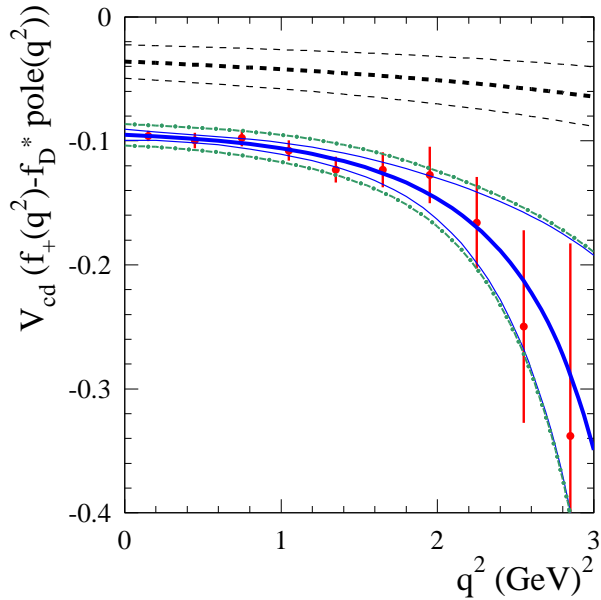}
  \end{center}
  \caption[]{Contributions of high-mass poles to $\Vcd \times f_{+,D}^{\pi}(q^2)$.
For all data and projections the $\Dstarp$ contributions are subtracted:
data points (red) represent the measurements (Fig. \ref{fig:babar_taylor}) and the full (blue) curve,
with thin lines on both sides, represents the fit result and uncertainties for the $z$-expansion parameterization, the dash-dotted lines indicate the additional uncertainties from the pole
estimate. 
The dashed black lines mark the expected contribution from the 
$D^{*\prime}_1$ pole (see Eq. (\ref{eq:res4})) and corresponding uncertainties. 
}
\label{fig:babar_cut}
\end{figure}

Figure\,\ref{fig:babar_cut} shows the difference between the present 
measurement, fitted with
the $z$-expansion parameterization, and the expectation
from the $\Dstarp$ pole contribution alone, as defined in Eq. (\ref{eq:res2}). 
On the same figure, the expected contribution from the first radial excitation
($D^{*\prime}_1$), as defined in Eq. (\ref{eq:res4}), is shown. 
This additional contribution cannot adequately describe the measurement.
The large difference between full and dashed lines illustrates the importance of contributions
from other hadronic states; the data do not favor a hadronic form 
factor ansatz with the  $D^*$ and $D^{*\prime}_1$ poles only.
 
\subsubsection{Test of the two-pole ansatz }
\label{sec:twopoles_physics}

The expected contribution of the $D^*$ pole, evaluated at $q^2=0$,
$Res(f^{\pi}_{+,D})_{D^*}/m^2_{D^*}=1.032\pm0.033$ 
(deduced from Eq. (\ref{eq:res2})) differs
from the value obtained with the modified-pole ansatz, 
$f^{\pi}_{+,D}(0)/(1-\alpha_{{\rm pole}})=0.85 \pm0.09$.
This indicates that the ansatz underestimates the $D^*$ pole contribution
by about two standard deviations.

While for the modified-pole ansatz an external condition
is used to eliminate one parameter, the two-pole ansatz has 
two parameters (see Eq. (\ref{eq:twopoles})). Data are fitted using 
this parameterization with the constraint
on the value of the residue expected for the $D^*$ pole (Eq. (\ref{eq:res2})).
The value, $\beta_{{\rm pole}}=0.68 \pm0.13\pm 0.11$, corresponds to an effective mass
for the second pole which is compatible with the $D^{*\prime}_1$ mass or
with an effective mass of several radial excitations,
\begin{equation}
m_{{\rm eff.}} = m_{D^*}/\sqrt{\beta_{{\rm pole}}}=\left ( 2.45^{+0.37}_{-0.26}\right )\,\gevcc,
\end{equation}
Presently, the measured contribution of the second pole, evaluated at $q^2=0$, is equal to
$-0.40 \pm 0.04 \pm 0.02$, which 
exceeds the expectation for the $D^{*\prime}_1$  of  $-0.16 \pm 0.06$, 
by a factor 2.5 (see  Eq. (\ref{eq:res4})).  
The parameter $\delta_{\rm pole}=0.47\pm0.21\pm0.18$ differs from zero,
the value expected in the modified-pole ansatz, by less than two standard 
deviations.

\subsubsection{Test of the three-pole ansatz}
\label{sec:threepoles_physics}
In fits to the fixed three-pole ansatz,
the residue for the second pole is constrained  to its
expected value.
The fitted value of the residue at the $D^*$ pole,
\begin{equation}
Res(f^{\pi}_{+,D})_{D^*}=(3.72\pm0.29\pm0.24) \,\gev^2,
\end{equation}
agrees to within one standard deviation with its expected value
(Eq. (\ref{eq:res2})).
This translates to the first experimental measurement of the $D^*$ decay
constant,
\begin{equation}
f_{D^*}= (219 \pm 17 \pm 14 )\,\mevcc.
\end{equation}
The value of the residue of the third pole is not accurately determined,
\begin{equation}
Res(f^{\pi}_{+,D})_{D^{*\prime}_2}=(-1.3\pm0.9\pm0.6) \,\gev^2.
\end{equation}

The sum of residues (see Eq. (\ref{eq:superconv})) is equal
to $(1.32 \pm 0.36 \pm 0.27)\,\gev^2$ and differs from zero by about three standard
deviations.
This result is obtained under the assumption that the third pole mass
 equals $3.1\,\gevcc$. On the other hand, 
several states above the $D^{*\prime}_1$ may contribute to an effective pole
at a higher mass. This possibility is
tested using the effective three-pole ansatz, by fitting the third pole mass, 
imposing the superconvergence 
condition and constraints on the first two residues
(see Fig. \ref{fig:babar_taylor}).  The fitted value of the effective pole mass is 
$m_{{\rm pole3}}=(3.6\pm0.3)\,\gevcc$, higher than the $D^{*\prime}_1$ mass, 
as expected. Fitted values of 
$Res(f^{\pi}_{+,D})_{D^*}=(4.12\pm0.13) \,\gev^2$
and  
$Res(f^{\pi}_{+,D})_{D^{*\prime}_1}=(-1.1\pm0.4) \,\gev^2$
are almost identical to the values used as constraints
(Eq. (\ref{eq:res2}-\ref{eq:res4})).
The ratio $\chi^2/\ndf=4.8/9$ indicates a good fit.
If the value of $\Vcd$ is allowed to vary in the fit,  the values of the
fitted parameters are
$\Vcd=0.20 \pm 0.02$ and $m_{{\rm pole3}}=(4.4\pm1.2)\,\gevcc$ for $\chi^2/\ndf=3.1/8$.

We conclude that the $q^2$ dependence  of the $\Ddecpi$ decay branching fraction   
is compatible with the effective three-pole ansatz for the form factor 
$f_{+,D}^{\pi}$ for which:
\begin{itemize}
\item the  values of the residues for the first two poles agree with
expectations;
\item the value of the third pole residue is obtained with
the superconvergence condition;
\item the third pole has an effective mass close to $4\,\gevcc$.  
\end{itemize}

\section{Extrapolation to $\Bdecpi$ decays}
\label{sec:dtob}

We implement two ways 
to use the information gained in this analysis of the $\Ddecpi$ decays
to extract a value for $\Vub$ from measurements of $\Bdecpi$ decays.

It is expected that lattice QCD calculations will eventually determine 
with high precision the ratio ($R_{BD}$) of the form factors
for charmless semileptonic decays of $B$ and $D$ mesons.
Until then, we have to rely on computations of the individual form factors 
(see Appendix \ref{sec:appendixc}) yielding an average value of
$R_{BD}=1.8 \pm 0.2$ for $w_H>4$, where $w_H$ is the product of the 
four-velocities of the heavy meson and the pion, defined in Appendix \
\ref{sec:appendixc}.
Based on Eq. (\ref{eq:ratio_width},\ref{eq:ratio_br}), 
the differential decay 
branching fraction for $\Bdecpi$ can be expressed as a function of $w=E^*_{\pi}/m_{\pi}$,
\begin{equation}
\frac{d {\cal B}^B}{d w} =\left .\frac{d {\cal B}^D}{d w}\right |_{{\rm meas.}} \frac{m_B~\tau_B}{m_D~\tau_D} \left (\frac{\Vub}{\Vcd} \right )^2 R_{BD}^2 ,
\label{eq:ratio_br2}
\end{equation}
here $E^*_{\pi}$ refers to the pion energy in the rest frame of the 
heavy meson (see Appendix \ref{sec:appendixc}).

Figure\,\ref{fig:bdinclexcl} compares the differential branching fraction 
$d{\cal B}^B/dw$ measured by \babar~\cite{ref:babarb} with the translated $\Ddecpi$ data, 
based on Eq.~(\ref{eq:ratio_br}), for $\Vub^{{\rm excl.}}=(3.23 \pm 0.31)\times 10^{-3}$,
the value extracted from $B \to \pi \ell \nu_{\ell}$ analyses~\cite{ref:pdg2013}.
In the common $w$ range, 
the two measured differential branching fractions are in good agreement,
probably not too surprising, since they are based on the same value of $\Vub$ 
and LQCD form factor normalizations. 
The result of the fit to the three-pole ansatz to the $\Ddecpi$ data 
is extrapolated into the unphysical region. 
The agreement with $\Bdecpi$
is good up to $w_B=11$ or $q^2>12 \gev^2$.
The fit based on the effective three-pole ansatz with
the superconvergence condition also describes the $\Bdecpi$ data
well, provided the ratio between the two form factors is independent of $w$.
The value of $\Vub$ obtained from a fit with this ansatz is,
\beq
\Vub = (3.65 \pm 0.18_{\rm exp.} \pm 0.40_{R_{BD}})\times 10^{-3}.
\label{eq:vub_ratio}
\eeq

\begin{figure}[!htb]
  \begin{center}
    \mbox{\epsfig{file=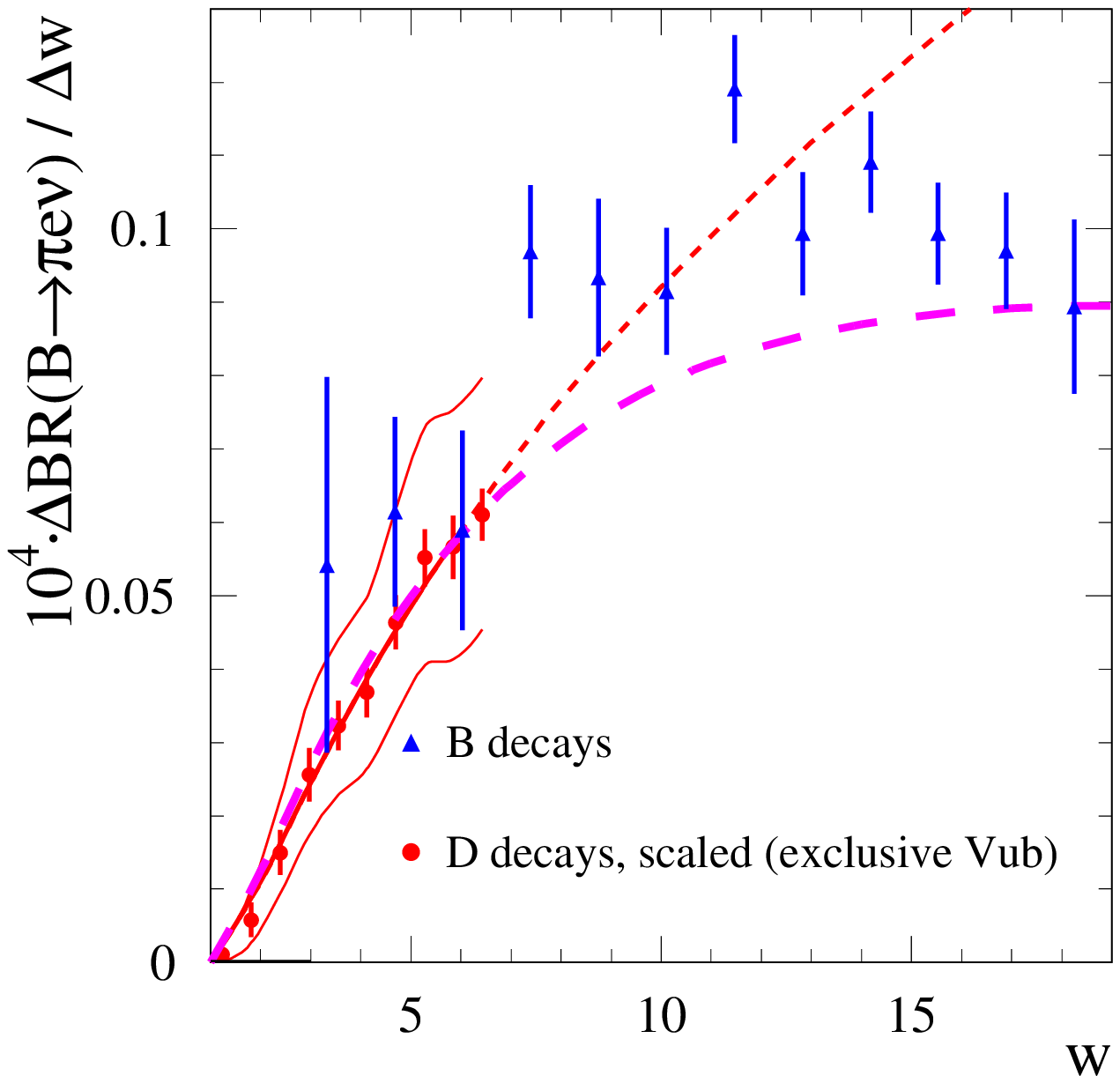,width=0.5\textwidth,angle=0}
}
  \end{center}
  \caption[]{ Comparison of the $d{\cal B}^B/dw$
differential decay rate for $\Bdecpi$ decays measured by \babar\ with an extrapolation 
of the $\Ddecpi$ form factor measurement.
The solid red line is the result of the fit to the fixed three-pole ansatz
with $m_{{\rm pole3}}=3.1\,\gevcc$, the short-dash red line marks the extrapolation
beyond the physical region for the $\Ddecpi$ decay. The two thin red lines indicate the impact of the 
12\% uncertainty on the form factor ratio $R_{BD}$.
The long-dash magenta line marks the fit result, to $\Ddecpi$ data, for the effective three-pole ansatz. In these comparisons, the value $\Vub^{{\rm excl.}}$ is used.}
  \label{fig:bdinclexcl}
\end{figure}

The second approach relies on the application of the effective three-pole ansatz
for $\Bdecpi$ decays,
\begin{eqnarray}
f_{+,B}^{\pi}(q^2) &=&Res(f_{+,B}^{\pi})_{B^*} \left ( \frac{1}{m_{B^*}^2-q^2} 
-\frac{d_2}{m_{B^{*\prime}_1}^2-q^2}\right . \nonumber\\
& &\left . -\frac{d_3}{m_{B^{*\prime}_2}^2-q^2} \right ),\, {\rm with} \, d_3=1-d_2 .
\label{eq:threepolesb}
\end{eqnarray}

\begin{figure}[!htb]
  \begin{center}
    \mbox{\epsfig{file=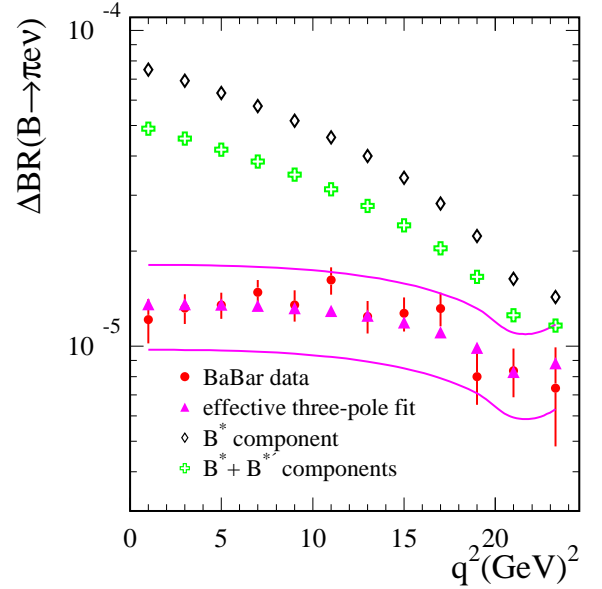,width=0.5\textwidth,angle=0}}
  \end{center}
  \caption[]{ Comparison of the measured differential branching fraction
for $\Bdecpi$~\cite{ref:babarb}, integrated over 2 $\GeV^2$ $q^2$ intervals
(apart for the last bin which extends from 22 to 26.2 $\GeV^2$),
with expectations from the effective three-pole ansatz. 
The two lines indicate theoretical uncertainties on these predictions.
The contributions from
the $B^*$ pole and from the sum of the $B^*$ and $B^{*\prime}_1$ poles
are indicated.}
  \label{fig:dbdq2_3poles}
\end{figure}
\noindent
It is expected that ratios of the residues at the different poles $d_2$ and $d_3$
are the same for $D$ and $B$ semileptonic decays~\cite{ref:burdman}.
Based on Eq. (\ref{eq:res2},\ref{eq:res4}),
we choose the value $d_2=0.26\pm0.10$.
For the $B^{*\prime}_1$ mass, we take $5.941\,\gevcc$ \cite{ref:defazio}
and the value of the third pole mass is a free parameter of the fit. 
The value of the residue of the form factor
at the $B^*$ pole is obtained from the following ratio,

\beq
\frac{Res(f_{+,B}^{\pi})_{B^*}}{Res(f_{+,D}^{\pi})_{D^*}}=
\frac{m^{3/2}_{B^*}m^{1/2}_{B}}{m^{3/2}_{D^*}m^{1/2}_{D}}
\frac{f_{B^*}}{f_B}\frac{f_{D}}{f_{D^*}}
\frac{\hat{g}_B}{\hat{g}_D}.
\label{eq:resb} 
\eeq
This expression is obtained from Eq. (\ref{eq:res1}) and the definition
of $\hat{g}_H$,
\beq
g_{H^*H\pi} = 2 \frac{\sqrt{m_{B^*}m_{B}}}{f_{\pi}}\hat{g}_H.
\label{eq:ghat} 
\eeq

The value of $\hat{g}_H$ is expected to be independent of the mass of the
heavy hadron. This has been verified, within present uncertainties, for $D$
and $B$ mesons: $\hat{g}_B=0.57\pm0.05\pm0.06$ \cite{ref:flynn},
$\hat{g}_D=0.53\pm0.03\pm0.03$ \cite{ref:becird}, and
$\hat{g}_{\infty}=0.52\pm0.05$ \cite{ref:ohki}  obtained
for an infinitely heavy hadron.
Based on recent LQCD calculations of the ratios of decay constants $f_{D^*}/f_D=1.20\pm0.02$
and $f_{B^*}/f_B=1.06\pm0.01$ \cite{ref:damirfdstar},  the
measured value of $f_D$, and the lattice result for $f_B=(190.5\pm4.2)\,\mev$
\cite{ref:flag}, we obtain,
\beq
\frac{Res(f_{+,B}^{\pi})_{B^*}}{Res(f_{+,D}^{\pi})_{D^*}}=
6.0 \pm 0.2 \pm 1.0
\label{eq:resb_over_resd}
\eeq
and 
\beq
Res(f_{+,B}^{\pi})_{B^*}=(24.9 \pm 1.2 \pm 4.0)\,\gev^2.
\label{eq:resbb}
\eeq
The second uncertainties in Eqs. (\ref{eq:resb_over_resd}-\ref{eq:resbb})
correspond to the uncertainty on the ratio $\hat{g}_B/\hat{g}_D$.

The expression in Eq. (\ref{eq:threepolesb}) is fitted to the \babar\
measurements of the $\Bdecpi$ decays~\cite{ref:babarb} with the 
residues at the two first poles, the effective mass of the third pole, and
$\Vub$ as free parameters. In addition, the residue at the $B^*$
pole must satisfy Eq. (\ref{eq:resbb}) and the value $d_2=0.26\pm0.10$
is constrained.

Figure\,\ref{fig:dbdq2_3poles} shows the result of the fit with 
$\chi^2/\ndf=10.7/10$. Fitted values of the quantities
entering in the constraints ($Res(f_{+,B}^{\pi})_{B^*}$ and $d_2$) 
and their corresponding uncertainties are almost identical to their
input values.
 Contributions of the $B^*$ pole alone
and of the two first poles are indicated. The $B^*$ pole component
is largely cancelled by hadronic states at higher masses.
The effective mass of the third pole is equal to $(7.4\pm0.4)\,\gevcc$.  
The fit results in
\beq
\Vub=(2.6 \pm 0.2_{\rm exp.} \pm0.4_{\rm theory})\times 10^{-3},
\eeq
a value that is compatible with the direct measurement based only on $\Bdecpi$
decays, using LQCD predictions for the form factor normalization. 
Here, the second uncertainty is related to the ratio $\hat{g}_B/\hat{g}_D$.
Other sources of systematic uncertainties are expected to be smaller:
\begin{itemize}
\item From the expected variation of the residues with the heavy quark mass, 
it is assumed that the ratio $d_2=0.26\pm0.1$ is the same for
$D$ and $B$ meson decays. A large change to $d_2= 0.5$ results in an increase 
in the value of $\Vub$ by $0.2\times 10^{-3}$, comparable to the measurement error;
\item The superconvergence condition~\cite{ref:burdman,ref:damirfdstar} is expected to be better satisfied for
$B$ than for $D$ decays because corrections in $1/m_H$ are smaller.
If we remove this condition from the fit and perform a scan as a 
function of the mass 
of the third pole, we observe that the superconvergence condition 
is satisfied for $m_{{\rm pole3}}<10 \,\gevcc$. Above this value, the residue
at the effective pole becomes large, but the fitted value of $\Vub$
decreases by only  $0.1\times 10^{-3}$, when $m_{{\rm pole3}}$ is varied
from $10 \,\gevcc$ to $100 \,\gevcc$.
\end{itemize}

\section{Summary}
\label{sec:Summary}

Based on a produced sample of 500 million $\ccbar$ events,
we have measured the ratio of the $\Ddecpi$ and $\Dz \rightarrow \Km \pip$
decay branching fractions,
\begin{equation}
 R_D  = 0.0702 \pm 0.0017 \pm 0.0023. \nonumber
\end{equation}
\noindent
Using the $\Dz \rightarrow \Km \pip$ branching fraction, given in 
Eq.~(\ref{eq:bkpidata}), we derive,
\begin{equation}
{\cal B}(\Dz \rightarrow \pim \ep \nue)= (2.770 \pm 0.068 \pm 0.092 \pm 0.037)\times 10^{-3},\nonumber
\end{equation}
where the third error accounts for the uncertainty 
on the branching fraction for the
$\Dz \rightarrow \Km \pip$ decay.

The measurements are sensitive to the product
$\Vcd \times f_{+,D}^{\pi}(q^2)$ and, using the $z$-expansion
parameterization of the hadronic form factor, we obtain:
\begin{equation}
\Vcd \times f_{+,D}^{\pi}(0)= 0.1374 \pm 0.0038 \pm 0.0022 \pm 0.0009 ,\nonumber
\end{equation}
where the last uncertainty corresponds to the uncertainties
on the branching fraction of the normalization channel and
on the $\Dz$ lifetime.
This measurement has an accuracy similar to previous
measurements by the CLEO-c Collaboration~\cite{ref:cleoc08,ref:cleoc09}.
 
We have measured the $q^2$ dependence of the differential branching fraction
(Table \ref{tab:errmeas}) and using the value of
$\Vcd=\Vus$, we have compared the $q^2$ variation of the hadronic
form factor with different parameterizations
(Table \ref{tab:fittedparam}). 

In general terms, the hadronic form factor can be expressed as an infinite
sum of pole contributions \cite{ref:burdman,ref:damirfdstar}.
At large $q^2$, the effective three-pole ansatz with the truncation of
the series to three poles, of which the third one is an effective pole, 
describes the measurements well, 
satisfying also the constraints from expectations for contributions of the first two poles.
This ansatz has been used to analyze $\Bdecpi$ decays and to
provide a parameterization for $f_{+,B}^{\pi}(q^2)$.
Using the fitted effective three-pole ansatz for $f_{+,D}^{\pi}(q^2)$
and assuming that the ratio $R_{BD}$ of the $B$ and $D$ form factors
does not depend on the pion energy, the value
$\Vub = (3.65 \pm 0.18 \pm0.40)\times 10^{-3}$ is obtained.
The dominant contribution to the systematic uncertainty originates from
$ R_{BD}$. 
In another approach, we have used the effective three-pole ansatz to fit the
measured partial branching fractions for the $\Bdecpi$
decays with constraints on  the value of the $B^*$
pole contribution and the ratio of the residues at the $B^{*\prime}_1$
and $B^*$ poles, taken to be equal to the corresponding ratio 
for charmed mesons \cite{ref:burdman,ref:damirfdstar}. We obtain
$\Vub = (2.6 \pm 0.2 \pm0.4)\times 10^{-3}$, where the dominant systematic
uncertainty originates from the residue at the $B^*$ pole.

These two values of $\Vub$ exploit common features of
$B$ and $D$ Cabibbo suppressed semileptonic decays,
as suggested many years ago, and should benefit from future improvements of the
measurements and of LQCD computations of the decay constants for charm and beauty mesons.

\vspace{0.5cm}
\section{Acknowledgments}
\label{sec:Acknowledgments}

The authors wish to thank D. Becirevic,
S. Descotes-Genon, and A. Le Yaouanc for their 
help with the theoretical interpretation of these results.

\input acknowledgements.tex

\appendix

\section{Values of parameters entering in the dispersive approach with constraints}
\label{sec:appendixb}

Using the expression for $Res(f_{+,D}^{\pi})_{D^*}$ in Eq. (\ref{eq:res1}),
$f_{D^*}/f_{D}=1.20 \pm 0.02 $, computed in LQCD 
\cite{ref:damirfdstar}, of $f_{D}=(204.4\pm5.0)\,\mev$ measured in experiments
\cite{ref:pdg2013},
and of $g_{D^{*+}D^0\pi^+}= 16.92 \pm0.13 \pm 0.14$
deduced
from the measurement of the intrinsic $D^{*+}$ width \cite{ref:babardstar},
the contribution of the $D^*$ pole in the $\Ddecpi$
decay channel is evaluated to be:
\beq
Res(f_{+,D}^{\pi})_{D^*}= (4.17 \pm 0.13)\,\gev^2.
\label{eq:res2} 
\eeq

In a similar way it is possible to evaluate the $D_1^{*\prime}$ contribution:
\beq
Res(f_{+,D}^{\pi})_{D_1^{*\prime}}=\frac{1}{2}m_{D_1^{*\prime}}\,f_{D_1^{*\prime}}\,g_{D_1^{*\prime}D^0\pi^+}
\label{eq:res3} 
\eeq
using the measured properties
of this first radial excitation  \cite{ref:dstarprime}  and taking, 
$f_{D_1^{*\prime}}= (148 \pm 45)\, \mev$, estimated from
a calculation of ratios of meson decay constants obtained in LQCD 
\cite{ref:damir2}.
The residue of the form factor at the
first radial excitation is then equal to:
\beq
Res(f_{+,D}^{\pi})_{D_1^{*\prime}}= (-1.1 \pm 0.4)\,\gev^2.
\label{eq:res4} 
\eeq
The negative sign is expected from LQCD evaluations 
 \cite{ref:dstarprimcoupling} and from 
phenomenological analyses \cite{ref:dstarwidthth}. The result we obtain
in this way agrees with values quoted in these references, which
were deduced under quite different assumptions. 
If measurements from LHCb  \cite{ref:dstarprime_lhcb} for the mass and width 
of the $D_1^{*\prime}$ meson are used in place of results from \babar\,
the central value of the residue estimate increases by $10\%$ and
this has no real effect on the present analysis, considering the
other sources of uncertainty.

The contribution from the $H\pi$ continuum with mass between threshold 
and the first radial excitation is evaluated in \cite{ref:burdman}
using chiral symmetry. Its importance is measured by the parameter:
\beq
c_H=\frac{1}{\pi} \int_{t_+}^{\Lambda^2} \mathcal{I}m(f_{+,H}^{\pi,{\rm cont.}}(t))dt
\label{eq:contnorm}
\eeq
Numerically we find that the continuum has a contribution of the order
of one third of that expected from the first radial excitation.
Following arguments in \cite{ref:damirfdstar}, it has been neglected. 

\section{ Relevant expressions in the $z$-expansion}
\label{sec:appendixa}

In terms of the variable $z$, the form factor, 
consistent with constraints from QCD, takes the form:
\beq
f_{+,D}^{\pi}(t)=\frac{1}{P(t) \Phi(t,t_0)}\sum_{k=0}^{\infty}a_k(t_0)~ z^k(t,t_0).
\eeq
This expansion in $z$ is expected to converge quickly.
The function
$P(t)$ accounts for the lowest mass pole at $t=m_{D^*}$, and
is equal to 1 because the pole is situated above the cut threshold; 
$\Phi$ is determined as
\beq
\Phi(t,t_0) &=& \sqrt{\frac{1}{24 \pi \chi_V}}
\left ( \frac{t_+-t}{t_+-t_0}\right )^{\frac{1}{4}}
\left ( \sqrt{t_+-t}+\sqrt{t_+} \right )^{-5}\nonumber\\
&\times&\left ( \sqrt{t_+-t}+\sqrt{t_+-t_0} \right )\\
&\times&\left ( \sqrt{t_+-t}+\sqrt{t_+-t_-} \right )^{\frac{3}{2}}\left ( t_+-t \right)^{\frac{3}{4}}\nonumber.
\eeq
\noindent 
The numerical factor $\chi_V$ can be calculated using 
perturbative QCD. It depends on $u=m_d/m_c$ \cite{ref:lebed}, and at leading order, with $u=0$, $\chi_V = 3/(32 \pi^2 m_c^2)$.
The functions $P(t)$ and $\Phi(t,t_0)$ are chosen such that
\beq
\sum_{k=0}^{\infty}a_k^2(t_0)\leq 1 \label{eq:unitary}.
\eeq
This constraint, which depends
on the choice of $\chi_V$, is not very useful for $D$ decays because the $\c$ quark mass is rather small and therefore may give rise to sizable $1/m_c$ and QCD corrections to $\chi_V$. 
However, the parameterization in Eq.~(\ref{eq:taylor}) remains valid and
it has been compared \cite{ref:hill1} with measurements, where
the first two terms in the expansion are sufficient to describe the data, 
given the current experimental uncertainties.

\section{Ratio between $B$ and $D$ form factors versus the pion energy}
\label{sec:appendixc}

Using the expression for the differential decay rate 
(see Eq. (\ref{eq:diff_decay_rate})), semileptonic branching
fractions for decaying mesons with different mass values are related.
Here, it is important to consider the decay rate for the 
same value of the energy of the emitted light meson ($E^*_{\pi}$), evaluated in the heavy meson rest frame.

The invariant-mass squared of the lepton system in terms of 
$E^*_{\pi}$ is equal to:
\begin{eqnarray}
q^2 = (p_H-p_{\pi})^2 &=& (E_H - E_{\pi})^2-(\vec{p}_H-\vec{p}_{\pi})^2\nonumber\\
& = & (m_H-E^*_{\pi})^2-\vec{p^*_{\pi}}^2\nonumber\\
& = & m_H^2 + m_{\pi}^2 -2~m_H E^*_{\pi}.
\end{eqnarray}

Instead of $E^*_{\pi}$, we can use the Lorentz invariant variable 
$w_H = v_H \cdot v_{\pi}$, where 
$v_{H}=p_{H}/m_{H}$ and $v_{\pi}=p_{\pi}/m_{\pi}$ 
are the 4-velocities of the
$H$ and $\pi$ mesons, respectively.
In terms of this quantity:
\beq
q^2 =  m_H^2 + m_{\pi}^2 -2~m_H m_{\pi}w_H.
\eeq

The differential semileptonic decay rate for a heavy meson 
($H$) versus $w_H$ is equal to:
\beq
\vspace{-1cm}
 \frac{d \Gamma^H}{d w_H}=-2 m_H m_{\pi}\frac{G^2_F}{24 \pi^3} \left | V_{hx} \right |^2
p_{\pi}^{*3} \left |f_{+,H}^{\pi}(w_H) \right |^2
\label{eq:diff_decay_rate2}
\eeq
in which the quantity $V_{hx}$ is the corresponding CKM matrix element.

At the same value of $w_H$, the pions emitted in the decay of two
heavy mesons with different mass values, have the same energy (and momentum).
It results that the ratio of the differential decay widths of the two heavy mesons can be written:
\beq
\frac{d \Gamma^B /d w_B }{d \Gamma^D /d w_D }= \frac{m_B}{m_D} \left (\frac{\Vub}{\Vcd} \right )^2 \left |\frac{f_{+,B}^{\pi}(w_B)}{f_{+,D}^{\pi}(w_D)}\right |^2.
\label{eq:ratio_width}
\eeq

In terms of differential branching fractions, the previous ratio is equal to:
\beq
\frac{d {\cal B}^B /d w_B }{d {\cal B}^D /d w_D }= \frac{\tau(B^0)}{\tau(D^0)}\frac{d \Gamma^B /d w_B }{d \Gamma^D /d w_D }
\label{eq:ratio_br}
\eeq

The minimum of the quantity $w_{B,D}~(=1)$ is obtained when the light meson and the
leptonic system are emitted at rest. This corresponds to the maximum 
$q^2_{H}$: $q^2_{H,{\rm max}}=(m_H-m_{\pi})^2$. Table \ref{tab:ranges}
lists the ranges spanned in the semileptonic 
decays of
$B$ and $D$ mesons in terms of $q^2$ and $w_{B,D}$ variables. 
The common interval in $w_{B,D}$ for $B$ and $D$ 
decaying to 
$\pi^- e^+ \nu_e$ is between $1$ and $6.72$ corresponding to the $q^2$ 
interval
$[26.4,~18]\,\gev^2$ for the $B$ meson decay. It is interesting to consider
the non-physical region of the $D$-meson decay, for negative
$q^2$ values. This is feasible if we have a parameterization for the form
factor $f_{+,D}^{\pi}(q^2)$ as, for example, the three-pole ansatz.

\begin{table}[!htbp!]
\begin{center}
 \caption[]{Ranges spanned by the $w_{B,D}$ and $q^2_{B,D}$ variables in $B$
and $D$ semileptonic decays where a pion is emitted.
  \label{tab:ranges}}
{\small
  \begin{tabular}{ccc}
    \hline\hline
$w_{B,D}$ & $q^2_B (\gev)^2$ & $q^2_D (\gev)^2$ \\
\hline
$1$ & $q^2_{B,{\rm max}}=26.42$ & $q^2_{D,{\rm max}}=2.98$\\
$w_{D,{\rm max}}=\frac{m_D^2+m_{\pi}^2}{2 m_D m_{\pi}}=6.72$ & $18.0$ & $0.0$\\
$w_{B,{\rm max}}=\frac{m_B^2+m_{\pi}^2}{2 m_B m_{\pi}}=18.93$ & $0.0$ & $-6.36$\\
    \hline\hline
  \end{tabular}
}

\end{center}

\end{table}
\begin{figure}[!htbp!]
  \begin{center}
\includegraphics[height=9cm]{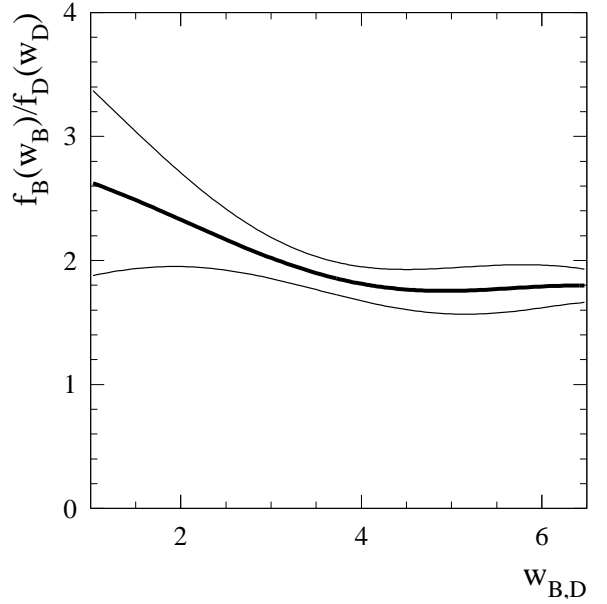}
  \end{center}
  \caption[]{    Variation of the ratio $f_{+,B}^{\pi}(w_B)/f_{+,D}^{\pi}(w_D)$ 
(thick line) computed from the evaluation of the two form factors 
obtained in LQCD \cite{ref:hpqcd3,ref:hpqcd2}. 
Thin lines give the uncertainties in this evaluation.}
 \label{fig:fB_fD_ratio}
\end{figure}

Based on the  scaling at large $q^2$ (close to $w_{B,D}=1$) we adopt the 
following approximation:
\beq
f_{+,H}^{\pi}(w_H) \sim \sqrt{m_H} \left [ f_{+,0}(w_H) +\frac{f_{+,1}(w_H)}{m_H}+... \right ].
\eeq
In this limit, the ratio between the $B$ and $D$ form factors is equal to,
\beq
\left |\frac{f_{+,B}^{\pi}(w_B)}{f_{+,D}^{\pi}(w_D)}\right |= \sqrt{\frac{m_B}{m_D}}
\left [ 1 + \mathcal{O} \left ( \frac{1}{m_{B,D}} \right )\right ],
\label{eq:fbfdratio}
\eeq
where the last term corresponds to neglected $1/m_H$ corrections.

The ratio of $f_{+,B}^{\pi}(q^2)$ \cite{ref:hpqcd3,ref:milcfnal1}
and  $f_{+,D}^{\pi}(q^2)$ \cite{ref:hpqcd2} values is shown in 
Fig.\,\ref{fig:fB_fD_ratio}.
It may be observed that:
\begin{itemize} 
\item the two form factors have a similar  $w$ dependence; 
\item for $w>4$, their ratio is 
 $1.8\pm0.2$. This value is not so different from the first order
expectation: $\sqrt{m_B/m_D}=1.7$;
\item the dependence of $f_{+,D}^{\pi}(w_D)$ and $f_{+,B}^{\pi}(w_B)$
on $w_D$ and  $w_B$, respectively, are very similar, thus their ratio
can be used to determine the absolute normalization
of the $B$ form-factor in this interval.
\end{itemize}

\section{Values of $\Vcd \times f_{+,D}^{\pi}(q^2)$
at the center of each bin}
\label{sec:appendixd}
We provide in Table \ref{tab:fq2_meas} the values displayed in Fig. \ref{fig:babar_taylor} of 
 $\Vcd \times f_{+,D}^{\pi}(q^2)$ evaluated at the center of each $q^2$
interval, for visual comparison with other measurements or theoretical expectations. 
Full uncertainty matrices are not provided because a detailed
numerical comparison with present measurements must use values given
in Table \ref{tab:errmeas} for the partial decay rates.

\begin{table*}[htbp]
  \caption {Values of 
 $\Vcd \times f_{+,D}^{\pi}(q^2)$ evaluated at the center of each $q^2$
interval with corresponding statistical and systematic uncertainties.
A  $0.7\%$ relative uncertainty, from the normalization channel and the $\Dz$
lifetime is common to all measurements and is not included.}
\begin{center}
\small{
  \begin{tabular}{lcccccccccc}
    \hline\hline
$q^2$ value $(\gev^2)$& 0.15 & 0.45 & 0.75 & 1.05  &1.35 & 1.65 & 1.95 & 2.25 & 2.55 & 2.85 \\ 
\hline
$\Vcd \times f_{+,D}^{\pi}(q^2)$ & 0.1455&  0.1620 & 0.1877&  0.2063&  0.2260 & 0.2697&  0.3219 & 0.3587&  0.3804&  0.4510  \\ 
stat. uncert. & 0.0037 &  0.0055 &  0.0060  & 0.0076  & 0.0098  & 0.0128  & 0.0182 &  0.0291 &  0.0693 &  0.1421 \\ 
syst. uncert. &  0.0019 &  0.0030 &  0.0028 &  0.0032 &  0.0038  & 0.0061 &  0.0138 &  0.0226 &  0.0349 &  0.0628 \\ 
    \hline\hline
  \end{tabular}}
\end{center}
  \label{tab:fq2_meas}
\end{table*}

\end{document}

%% file: authors_jun2014.tex
%
\author{J.~P.~Lees}
\author{V.~Poireau}
\author{V.~Tisserand}
\affiliation{Laboratoire d'Annecy-le-Vieux de Physique des Particules (LAPP), Universit\'e de Savoie, CNRS/IN2P3,  F-74941 Annecy-Le-Vieux, France}
\author{E.~Grauges}
\affiliation{Universitat de Barcelona, Facultat de Fisica, Departament ECM, E-08028 Barcelona, Spain }
\author{A.~Palano$^{ab}$ }
\affiliation{INFN Sezione di Bari$^{a}$; Dipartimento di Fisica, Universit\`a di Bari$^{b}$, I-70126 Bari, Italy }
\author{G.~Eigen}
\author{B.~Stugu}
\affiliation{University of Bergen, Institute of Physics, N-5007 Bergen, Norway }
\author{D.~N.~Brown}
\author{L.~T.~Kerth}
\author{Yu.~G.~Kolomensky}
\author{M.~J.~Lee}
\author{G.~Lynch}
\affiliation{Lawrence Berkeley National Laboratory and University of California, Berkeley, California 94720, USA }
% ...
\collaboration{The \babar\ Collaboration}
\noaffiliation

%% file: acknowledgements.tex
We are grateful for the 
extraordinary contributions of our \pep2\ colleagues in
achieving the excellent luminosity and machine conditions
that have made this work possible.
The success of this project also relies critically on the 
expertise and dedication of the computing organizations that 
support \babar.
The collaborating institutions wish to thank 
SLAC for its support and the kind hospitality extended to them. 
This work is supported by the
US Department of Energy
and National Science Foundation, the
Natural Sciences and Engineering Research Council (Canada),
the Commissariat \`a l'Energie Atomique and
Institut National de Physique Nucl\'eaire et de Physique des Particules
(France), the
Bundesministerium f\"ur Bildung und Forschung and
Deutsche Forschungsgemeinschaft
(Germany), the
Istituto Nazionale di Fisica Nucleare (Italy),
the Foundation for Fundamental Research on Matter (The Netherlands),
the Research Council of Norway, the
Ministry of Education and Science of the Russian Federation, 
Ministerio de Econom\'{\i}a y Competitividad (Spain), the
Science and Technology Facilities Council (United Kingdom),
and the Binational Science Foundation (U.S.-Israel).
Individuals have received support from 
the Marie-Curie IEF program (European Union) and the A. P. Sloan Foundation (USA). 
